\documentclass[cits,pdftex]{PoS}
\usepackage{graphicx,wrapfig}
\usepackage{macros}
\usepackage{multirow}
\usepackage{amssymb,amsmath}
\usepackage{booktabs,dcolumn}
\usepackage{grffile}         
\newcolumntype{C}{>{$}c<{$}} 
\title{%
Charmed pseudoscalar decay constants on\\ 
three-flavour CLS ensembles with open boundaries
}

\author{%
\includegraphics[width=2.125cm]{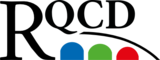}%
\hfill%
\includegraphics[width=2.5cm]{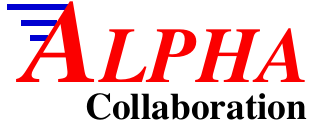}%

\hfill\parbox{17.5mm}{\vspace{0.25cm}\raggedleft\fns\it%
MS-TP-16-36
}
\vspace{0.25cm}
}

\ShortTitle{
Charmed pseudoscalar decay constants on $\nf=2+1$ CLS ensembles
with open boundaries
}

\author{%
Sara Collins$\,^a$,
Kevin Eckert\thanks{Speaker.}$\;\;^b$,
Jochen Heitger$\,^b$,
Stefan Hofmann\thanks{Speaker.}$\;\;^a$
and Wolfgang S\"oldner$\,^a$\\\\
{\rm (RQCD and ALPHA Collaborations)}\\\\
\llap{$^a$}%
Universit\"at Regensburg,
Institut f\"ur Theoretische Physik,
D-93040 Regensburg, Germany\\
E-mail: \email{sara.collins@ur.de,stefan.hofmann@ur.de,wolfgang.soeldner@ur.de}\\\\
\llap{$^b$}%
Westf\"alische Wilhelms-Universit\"at M\"unster, 
Institut f\"ur Theoretische Physik,\\
Wilhelm-Klemm-Stra{\ss}e 9, D-48149 M\"unster, Germany\\
E-mail: \email{k\_ecke03@uni-muenster.de,heitger@uni-muenster.de}
}

\abstract{%
\noindent
We determine the masses and pseudoscalar decay constants of D and 
$\mathrm{D}_\mathrm{s}$ mesons employing lattice QCD with 
non-perturbatively ${\cal O}(a)$ improved Wilson quarks and a tree-level 
Symanzik-improved gauge action.
Our analysis is based on the large-volume $\nf=2+1$ ensembles using open
boundary conditions, generated within the CLS effort. 
The status of results presented here covers two lattice spacings,
$a\approx 0.0854$ fm and $a\approx 0.0644$ fm, and pion masses varied
from 420 to 200 MeV.
We also report on our implementation of distance
preconditioning for the calculation of heavy quark propagators and discuss
the impact of the resulting accuracy improvements on the extraction of
charmed meson masses and decay constants.
This is part of a continuing analysis by the RQCD and ALPHA Collaborations,
aiming at a stable continuum extrapolation using several lattice spacings.
To extrapolate to the physical masses, we follow both,
the \mbox{$(2\ml+\ms)=\mathrm{const.}$} and the $\ms=\mathrm{const.}$ line
in parameter space.

\hfill\includegraphics[width=1.5cm]{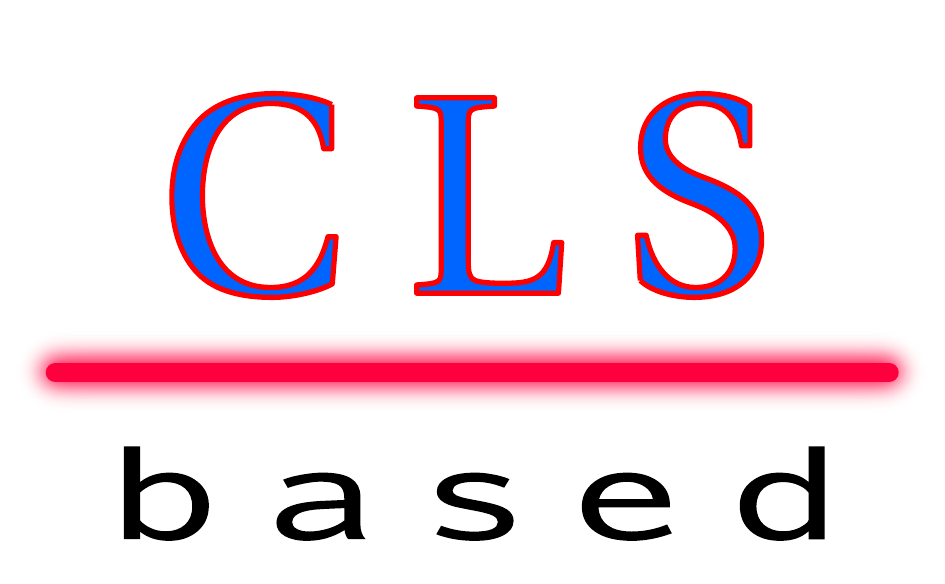}
}

\FullConference{%
34th annual International Symposium on Lattice Field Theory\\
24-30 July 2016\\
University of Southampton, UK}
\begin{document}
%
%
\section{Introduction}
\label{sec:intro}
\noindent
Recent experiments at the LHC, BEPC II and KEK supply us with an
abundance of spectroscopic information about charmed systems with ever
improving measurements of the leptonic decay rates of D and
$\mathrm{D}_\mathrm{s}$ mesons.  In order to fully exploit this
progress on the experimental side and to convert it into an amended
determination of the CKM matrix entries $\big|V_{\rm cq}\big|$,
${\rm q}={\rm d},{\rm s}$, it must be complemented on the lattice QCD side
with a more precise computation of the associated low-energy hadronic
matrix elements, characterised by the pseudoscalar decay constants
$\fD$ and $\fDs$, respectively.  Charmed systems are challenging
because of systematics, in particular, cutoff effects are generally
significant.  Therefore, all the more important is a fully controlled
continuum extrapolation, $a \rightarrow 0$.  In this proceedings we
describe our ongoing efforts in calculating the charmed pseudoscalar decay
constants.

\section{CLS ensembles and general computational setup}
\label{sec:cls}
\noindent
Our analysis is based on the $\nf=2+1$ Coordinated Lattice Simulations
(CLS) ensembles generated with non-perturbatively ${\cal O}(a)$
improved Wilson-Sheikholeslami-Wohlert (clover) fermions and the
tree-level improved L\"uscher-Weisz gauge action, employing the
open-source package \texttt{openQCD}~\cite{openQCD}.
  \begin{figure}[b!]
    \centering
    \includegraphics[width=0.8\textwidth]{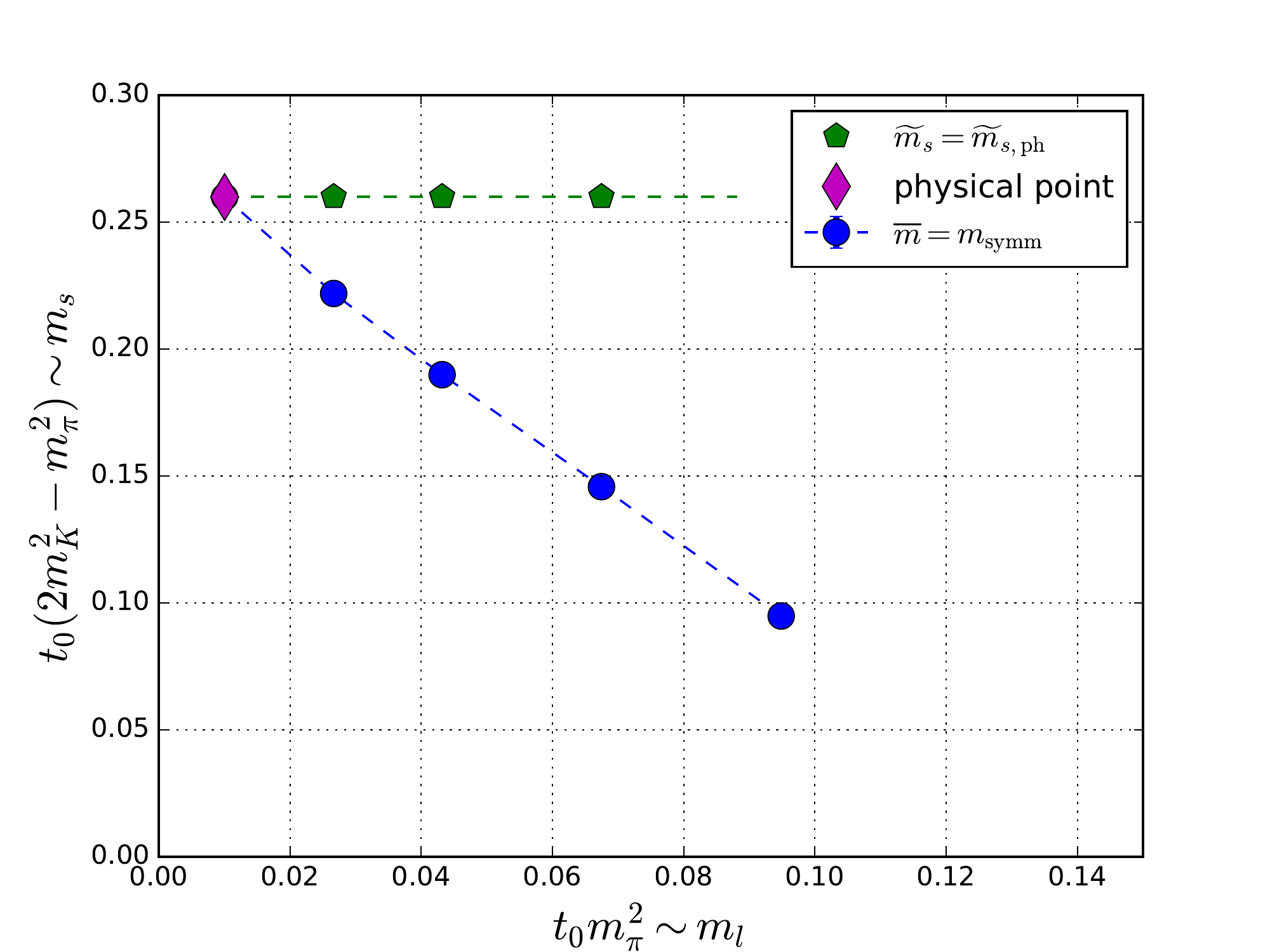}
    \caption{The light and strange quark masses realised for 
      $\beta=3.4$ lattices as indicated by the square of the pion mass
      versus the kaon-pion mass difference,
      $2m_K^2-m_\pi^2$, in units of $t_0$.}
    \label{fig:QuarkMassPlane}
  \end{figure}
Ensembles have
been realised with lattice spacings in the range of 
$a \approx 0.0854-0.039\,\Fm$ ($\beta=3.4-3.85$), aiming at a
controlled extrapolation to the continuum
limit~\cite{Bruno:2014jqa,Bali:2016umi,Bruno:2016plf}.  Simulating
with $a <0.05$ fm is possible through the use of open boundary
conditions in the temporal direction, which avoids the problem of
topological freezing as the continuum limit is approached
\cite{Luscher:2011kk,Luscher:2012av}.  Readers, who are interested in
the details regarding the algorithmic setup, are referred to
Ref.~\cite{Bruno:2014jqa}.  Here, we only mention the use of
twisted-mass reweighting for the two mass-degenerate light
fermions~\cite{Luscher:2008tw}, which prevents instabilities in the
HMC algorithm due to near-zero modes of the Dirac operator, and the
use of the rational approximation for simulating the strange
quark. These algorithmic choices mean that the computation of
observables, such as correlation functions, require reweighting, see
Section~\ref{sec:ana}.

\begin{table}
{
\footnotesize
\begin{tabular}{ccllcccccc}
\toprule
trajectory&ensemble&\multicolumn{1}{c}{$\kapl$}&\multicolumn{1}{c}{$\kaps$}&$Lm_{\pi}$&$N_t\times N_s^3$&$\frac{m_{\pi}}{\textmd{MeV}}$&
$\frac{m_K}{\textmd{MeV}}$&$N_{\mathrm{MD}}$\\
\midrule
\multicolumn{9}{c}{$\beta=3.4$ [$a= 0.0854(15)\,\textmd{fm}$], \hspace{5pt} $\sqrt{8t_0}/a=4.852(7)$}\\
\midrule
\multirow{4}{*}{$\overline{m}=m_{\mathrm{sym}}$}
  &H101   &0.13675962 &0.13675962    & 5.8 & $96\times 32^3$  & 
  422 & 422 &  8000\\
  &H102   &0.136865   &0.136549339   & 4.9 & $96\times 32^3$  &
  356 & 442 &  7988\\
  &H105   &0.13697    &0.13634079    & 3.9 & $96\times 32^3$  &
  282 & 467 &  11332\\
  &C101   &0.13703    &0.136222041   & 4.6 & $96\times 48^3$  &
  223 & 476 &  6208\\
\hline
\multirow{3}{*}{$\hat{m}_{\rm s}=\hat{m}_{\rm s}^{\rm phys}$}
  &H107&0.136945665908&0.136203165143 & 5.1 & $96\times 32^3$&
  368&549&6256\\
  &H106&0.137015570024&0.136148704478 & 3.8 & $96\times 32^3$&
  272&519&6212\\
  &C102&0.1370508458  &0.136129062556 & 4.6 & $96\times 48^3$&
  223&504&6000\\
\midrule
\multicolumn{9}{c}{$\beta=3.55$ [$a= 0.0644(11)\,\textmd{fm}$], \hspace{5pt} $\sqrt{8t_0}/a=6.433(6)$}\\
\midrule
\multirow{4}{*}{$\overline{m}=m_{\mathrm{sym}}$}
  & H200 & 0.137    & 0.137       & 4.4& $96\times 32^3$ &
  418 & 418 & 8000\\
  & N203 & 0.13708  & 0.136840284 & 5.4& $128\times 48^3$ &
  345 & 441 & 6172\\
  & N200 & 0.13714  & 0.13672086  & 4.4& $128\times 48^3$ &
  283 & 461 & 6800\\
  & D200 & 0.1372   & 0.136601748 & 4.2& $128\times 64^3$&
  199 & 479 & 4000\\
\hline
\multirow{3}{*}{$\hat{m}_{\rm s}=\hat{m}_{\rm s}^{\rm phys}$}
  & N204 & 0.137112   & 0.136575049 & 5.6& $128\times 48^3$&
  351 & 544 & 2000\\
  & N201 & 0.13715968 & 0.136561319 & 4.5& $128\times 48^3$&
  284 & 522 & 6000\\
  & D201 & 0.137207   & 0.136546436 & 4.1& $128\times 64^3$&
  198 & 499 & 4312\\
\bottomrule
\end{tabular} }
\caption{Details of the ensembles analysed so far for the two
  trajectories to the physical point, keeping $\overline{m}$ fixed to
  the value at the symmetric point~($\overline{m}=m_{\rm sym}$) and
  keeping the renormalised strange quark mass~($\hat{m}_{\rm s}$) approximately equal to
  the physical value~($\hat{m}_{\rm s}^{\rm phys}$). The light and strange
  quark hopping parameters are denoted $\kapl$ and $\kaps$,
  respectively. The lattice volumes $N_t\times N_s^3$,
  the pion~($m_\pi$) and kaon~($m_K$) masses and the statistics given by
  the number of molecular dynamics units~($N_{\rm MD}$) are also indicated.}
\label{tab:CLSensembles}
\end{table}

So far we have analysed lattices with $\beta=3.4$ and $3.55$
corresponding to $a=0.0854(15)$ and $0.0644(11)\,\Fm$, 
respectively~\cite{Bali:2016umi}.
For each $\beta$-value, the physical point is approached in the light and
strange quark mass plane following 2 trajectories --- (i) keeping the
average lattice quark
mass~($\overline{m}=\left(2\ml+\ms\right)/3$~\footnote{%
Here we refer to the vector Ward identity masses, 
$m_{{\rm q}={\rm l(ight)},{\rm s}}=
(1/\kappa_{\rm q}-1/\kappa_{\rm crit})/(2a)$,
where $\kappa_{\rm crit}$ is the critical hopping parameter value
at which the axial Ward identity mass in the symmetric limit, $\ml=\ms$,
vanishes.})
fixed such that the sum of the renormalised quark masses is constant up
to ${\cal O}(a)$ effects, and (ii) keeping the renormalised strange
quark mass approximately constant. Figure~\ref{fig:QuarkMassPlane}
displays the two trajectories for $\beta=3.4$, where the pion masses
are varied from 422 down to 223 MeV. Simulations along the
$\overline{m}$ constant line, first proposed by the
QCDSF Collaboration~\cite{Bietenholz:2010jr}, start from the
$\ml=\ms=m_{\rm sym}$ symmetric point with the strange~(light) quark mass 
becoming heavier~(lighter) towards the physical point.
The quark mass dependence of hadron observables is given by Gell-Mann-Okubo
expansions in powers of $\ms - \ml$ or, alternatively, can be described
by SU(3) chiral perturbation theory~(ChPT). For the second trajectory,
SU(2) ChPT applies. See Ref.~\cite{Bali:2016umi} for details on how
an almost constant renormalised strange quark mass was achieved in
this case.
As will be demonstrated in Section~\ref{sec:res}, having two
trajectories available enables the extrapolation to the physical point
to be tightly constrained. 

Table \ref{tab:CLSensembles} provides further details of the ensembles
entering our analysis.  They exhibit large spatial volumes with high
statistics; in particular, $m_\pi L\gtrsim4$ ($L=N_sa$) is satisfied by all
ensembles, and more than 4000 molecular dynamic units
$N_{\mathrm{MD}}$ have been generated in each case~(with the exception
of ensemble N204, which is still in production).  For open temporal
boundaries, one needs large physical time extents, since
discretisation effects and the propagation of scalar particles into
the bulk mean that, in general, timeslices close to the boundaries
need to be discarded. The lattice spacing is set by the gradient
(resp. Wilson) flow scale $t_0$, where we use 
$\sqrt{8t_0^{\rm phys}}=0.4144(59)(37)\mathrm{fm}$, as determined by BMW
Collaboration~\cite{Borsanyi:2012zs}, together with
$\sqrt{8t_0}/a$-values from Refs.~\cite{Bruno:2014jqa,Bali:2016umi}.
Using the pion and kaon decay constants to set the scale, is
discussed in Ref.~\cite{Bruno:2016plf}.  The physical point is defined
employing the combinations \be \phi_4=
8t_0\left(m_K^2+\frac{1}{2}\,m_\pi^2\right) \qquad\mathrm{and}\qquad
\phi_2= 8t_0m_\pi^2, \ee together with the pion and kaon masses in the
isospin limit, $m_\pi=134.8(3)\,\MeV$ and $m_K=494.2(4)$ MeV, taken
from the FLAG report~\cite{Aoki:2016frl}. We note that the starting
value for the $\overline{m}=\mathrm{const.}$ trajectory at the
symmetric point, $m_{\rm sym}$, was fixed by tuning $\phi_4=1.15$,
slightly above the physical value of $\phi_4^{\rm phys}=1.117(38)$.
This choice was motivated by expected discretisation and ${\cal
  O}\big((\ms - \ml)^2\big)$ effects in $\phi_4$ along the
trajectory. In addition, the computation of $t_0$ involved in this
tuning procedure was performed at unphysical quark mass.  In the
final analysis, the trajectory may be found to slightly miss the
physical point.  Mistunings of this kind, however, can still be
accounted for later by slightly adjusting the results through
reweighting or via a Taylor expansion~(following
Ref.~\cite{Bruno:2016plf}).  We leave these considerations for future
study.

\section{Definition of observables}
\label{sec:obs}
\noindent
Let us explain our notation and the observables analysed.  The
pseudoscalar decay constants $\fD$ and $\fDs$ are defined via the
matrix elements of the axial vector current between $\mathrm{D}$ and
$\mathrm{D}_\mathrm{s}$ meson states $|\mathrm{D}(p)\rangle$ and
$|\mathrm{D}_\mathrm{s}(p)\rangle$ at momentum $p$, respectively, and
the vacuum, viz.
\begin{equation}
 \langle 0\left|A^{\rm lc}_\mu\right|\mathrm{D}(p)\rangle=
 \mathrm{i}\fD p_\mu,
 \qquad  
 \langle 0\left|A^{\rm sc}_\mu\right|\mathrm{D}_\mathrm{s}(p)\rangle=
 \mathrm{i}\fDs p_\mu,
 \label{eq:Fps}
\end{equation}
where
$A^{\rm qc}_{\mu}=\overline{q}\gamma_\mu\gamma_5c$ for quark flavours
${\rm q}={\rm l(ight)},{\rm s}$.  
In order to reduce leading discretisation effects to ${\cal O}(a^2)$ in the
lattice simulation, we employ the improved axial operator
\begin{equation}
 A^{\rm qc,I}_{\mu}= 
 A^{\rm qc}_{\mu}
 +a\ca\frac{1}{2}\left(\partial_\mu+\partial^*_\mu\right)P^{\rm qc}.
 \label{eq:CaImprove}
\end{equation}
The pseudoscalar interpolator is given by
$P^{\rm qc}=\overline{q}\gamma_5c$. The improvement coefficient $\ca$ has
been determined non-perturbatively in Ref.~\cite{Bulava:2015bxa}. 
Due to the breaking of chiral symmetry by Wilson-type
fermions, the axial current must be renormalised in order to ensure that the
continuum axial Ward identity
\begin{equation}
 \left(A^{qc}_{\mu}\right)^{\rm R}=
 \za\left[1+a\left(\ba m_{\rm qc}+3\tilde{b}_{\rm A}\overline{m}\right)\right]
 A^{\rm qc,I}_{\mu} +{\cal O}(a^2)
 \label{eq:RenImp}
\end{equation}
is satisfied, where the (bare) vector Ward identity quark mass combinations 
read:
\begin{equation}
 m_{\rm qc}=
 \frac{1}{2}\left(\mq+\mc\right), \qquad 3\overline{m}=\ms+2\ml.
\end{equation}
The renormalisation factor $\za$ is taken from the non-perturbative
evaluation of Ref.~\cite{Bulava:2016ktf}. The coefficient $\ba$ has
been determined non-perturbatively in Ref.~\cite{Korcyl:2016ugy}.
The same authors find $\tilde{b}_{\rm A}$ to be consistent with zero in a
preliminary analysis at $\beta=3.4$.
In order to perform a consistent analysis for both $\beta=3.4$ and 3.55,
at present, we neglect the $\tilde{b}_{\rm A}\overline{m}$-term.
Since the mass dependent corrections in Eq.~(\ref{eq:RenImp}) are
anyway dominated by the term involving $m_{\rm qc}$, this choice does not
introduce a significant systematic uncertainty.

In the following, we restrict the analysis to zero-momentum 
$\mathrm{D}$ and $\mathrm{D}_s$ mesons, for which only the temporal
component of the axial current needs to be considered. The matrix elements
in Eq.~(\ref{eq:Fps}) are extracted from the following two-point
correlation functions:
\begin{eqnarray}
 C_{\rm A}(x_0, y_0)=
 -\frac{a^6}{L^3}\sum_{\vec{x},\vec{y}}
 \langle A_4^{\rm qc,I}(x)\left(P^{\rm qc}(y)\right)^\dagger\rangle, \qquad
 C_{\rm P}(x_0, y_0)=
 -\frac{a^6}{L^3}\sum_{\vec{x},\vec{y}}
 \langle P^{\rm qc}(x)\left(P^{\rm qc}(y)\right)^\dagger\rangle,
 \label{eq:corrs}
\end{eqnarray}
with the source and sink interpolating operators inserted at timeslices 
$y_0$ and $x_0$, respectively. 
For sufficiently large time differences $x_0-y_0$ and $T-x_0$, for a
lattice of physical time 
extent~\footnote{%
For lattices with $N_t$ points in temporal direction, $T=(N_t-1)a$
denotes the physical time extent.} $T$, these correlators
behave as~\cite{Bruno:2016plf} 
\begin{eqnarray}
\label{eq:corrs_exp}
 C_{\rm A}(x_0,y_0)\approx 
 \frac{f^{\rm bare}_{\rm qc}}{2}\,A(y_0)\,\mathrm{e}^{-m_{D_{\rm q}}(x_0-y_0)},
 \qquad
 C_{\rm P}(x_0,y_0)\approx 
 \frac{\left|A(y_0)\right|^2}{2m_{D_{\rm q}}}\,
 \mathrm{e}^{-m_{D_{\rm q}}\left(x_0-y_0\right)},
\end{eqnarray}
where $m_{D_{\rm q}}$ is the mass of a pseudoscalar meson containing a 
charm quark and an anti-quark of flavour ${\rm q}={\rm l},{\rm s}$. 
The (unrenormalised) lattice decay constant, $f^{\rm bare}_{\rm qc}$,
corresponds to $\langle 0|A_4^{\rm qc,I}|\mathrm{D}_{\rm q}\rangle/m_{D_{\rm q}}$,
while the source-time dependent amplitude $A(y_0)$ encapsulates the matrix 
element of the pseudoscalar operator 
$\langle 0\left|P^{\rm qc}\right|\mathrm{D}_{\rm q}\rangle$ and 
effects arising from the boundary close to the source.
For $y_0$ sufficiently far from the boundary,
$A(y_0)=\langle 0\left|P^{\rm qc}\right|\mathrm{D}_{\rm q}\rangle$ holds.

\begin{figure}[b!]
 \centering
 \includegraphics[width=0.5\textwidth]{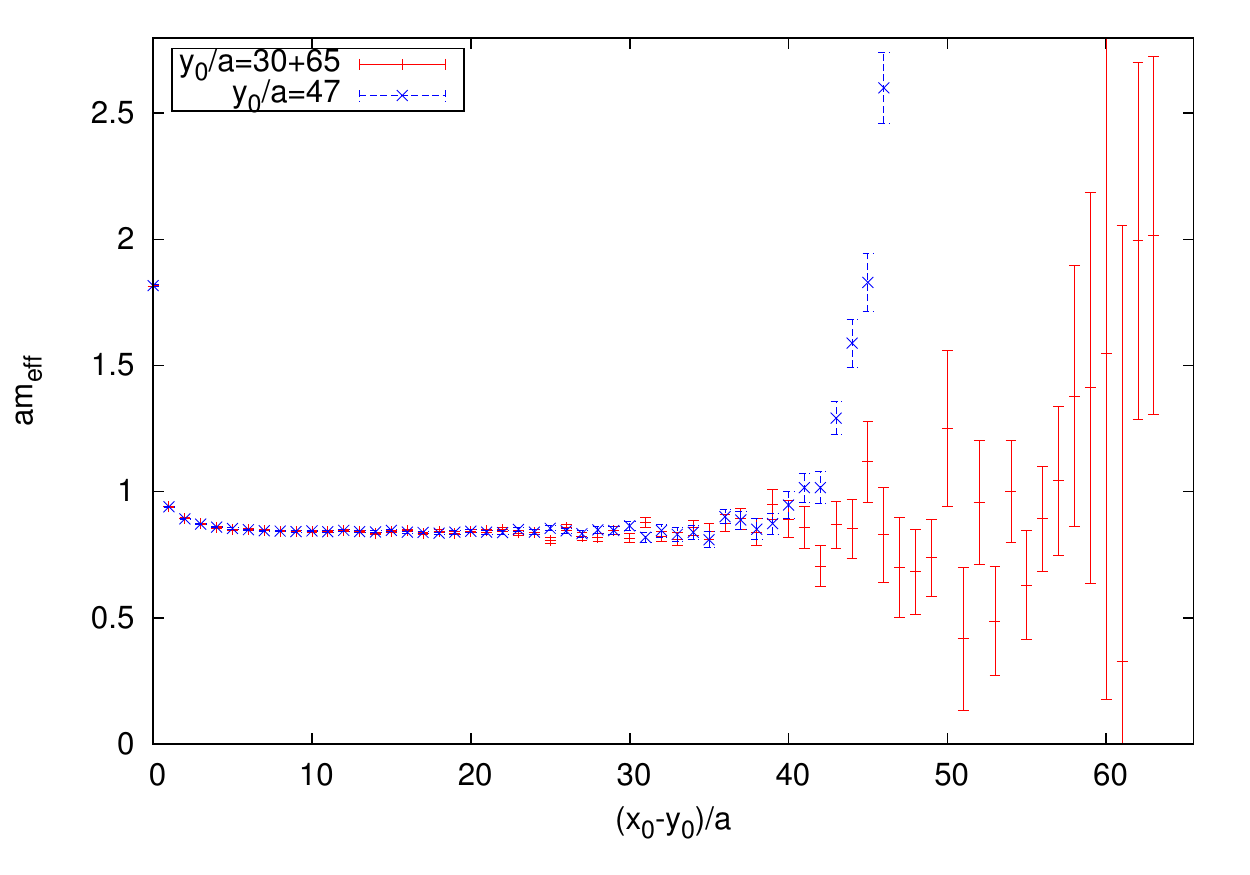}
 \caption{The effective mass of the pseudoscalar D meson as a function
   of $(x_0-y_0)/a$ for ensemble H105 with $T/a=95$.  Three source
   positions were used: $y_0/a=30$, 47 and 65.  The correlation
   function of the forward-propagating meson for $y_0=30a$ is averaged
   with that of the backward-propagating meson for $y_0=65a$, yielding
   the effective mass indicated by the red ``+'' symbols.  For
   $y_0/a=47$, the forward- and backward-propagating mesons are
   averaged to give $am_{\rm eff}$ shown as the blue crosses. }
 \label{fig:effMassDiffSource}
\end{figure}
In this preliminary analysis we place the source at large enough times
such that boundary effects can be neglected. In order to gain
statistics, three sources are chosen corresponding to $y_0/a=30$, 47
and 65 for ensembles with $T/a=95$ and $\beta=3.4$~(see
Table~\ref{tab:CLSensembles}), while for $\beta=3.55$, the sources are
placed at $y_0/a=27$, 47 and 68 for ensembles with $T/a=95$, and at
$y_0/a=46$, 63 and 81 for those with $T/a=127$. Ground-state dominance
at small source-sink separations is realised by employing Wuppertal
smearing~\cite{Gusken:1989ad,Gusken:1989qx} with APE-smoothed
links~\cite{Falcioni:1984ei} to the pseudoscalar source and sink
interpolators. Figure~\ref{fig:effMassDiffSource} displays typical
effective masses for the D meson derived from the correlation function
$C_{\rm P}(x_0,y_0)$ for the different source positions as a function
of the sink-source time separation. For small $x_0-y_0$, the results
from the source in the center of the lattice~($y_0/a=47$) are
compatible with those from the sources closer to the boundary~(i.e.,
from the forward-propagating meson for $y_0=30a$ and the
backward-propagating meson for $y_0=65a$), consistent with the absence
of boundary effects.  However, the latter are clearly evident for
$y_0=47a$, as the sink time nears the boundary, while
for $y_0=30a$ and 65$a$ the signal deteriorates before significant
boundary effects are seen.

The pseudoscalar decay constant is extracted by fitting 
$C_{\rm A}(x_0,y_0)$ and $C_{\rm P}(x_0,y_0)$ to the functional forms given
in Eq.~(\ref{eq:corrs_exp}), with $A(y_0)$ independent of $y_0$.
The fitting procedure is discussed in Section~\ref{sec:ana}.
Alternative approaches, for instance, employing the axial Ward identity 
(i.e., PCAC) quark mass~\cite{Heitger:2013oaa}, will also be considered in
the future.
We remark that for pseudoscalar mesons containing a quark and an
anti-quark of the same flavour, for which the statistical noise of the
correlators is constant with varying source-sink separation, one can also
place the source close to the boundary and consider ratios such as
\begin{equation} 
 f^{\rm bare}_{\rm qq}(x_0,y_0)\propto 
\sqrt{\frac{C_{\rm A}(x_0,y_0)C_{\rm A}(x_0,T-y_0)}{C_{\rm P}(T-y_0,y_0)}},
\end{equation}
see Ref.~\cite{Bruno:2014lra} for details.
When extracting the decay constant from the correlation functions, 
we wish to maximise the time range for the fit. However, there is a well 
known problem with numerical accuracy towards large time separations,
when computing correlation functions involving propagators of heavy quarks. 
This issue is discussed in the next section along with the distance
preconditioning procedure of Ref.~\cite{deDivitiis:2010ya}, which can be
adopted to alleviate such difficulties.

\section{Distance preconditioning of heavy quark propagators}
\label{sec:dp}
\noindent
\subsection{Motivation for distance preconditioning}
\noindent
When numerically inverting the Dirac operator, one has to pay
special attention to the quality of the obtained solutions. 
Internally, the solver routine checks if the condition
\begin{equation}
\Big|\sum_y (D[U]+m_0)_{x,y}\,S^n (y)-\eta_t(x)\Big|
< 
r_{\text{gl}}{}
\label{eq:residual}
\end{equation}
is satisfied, where $D[U]$ is the discretized lattice Dirac operator, 
$m_0$ the bare quark mass in lattice units and $S^n(y)$ the approximate 
solution at the $n$-th iteration of the solver.
$\eta_t(x)$ denotes a source located on a single timeslice 
$t$, while $r_{\text{gl}}$ is the global residuum, i.e., indicating the 
numerical accuracy one likes to achieve. 
Contributions to the norm above are negligible for heavy quarks, 
if one considers large time separations between the source $y_0$ and the
sink $x_0$, because of the exponential decay of correlation functions
$\propto \exp(-m_0 \vert y_0 - x_0 \vert)$. 
Solutions for large time extents thus become increasingly inaccurate in 
this case.

In order to improve the overall precision in our charm quark propagator
computations, we have implemented the so-called
``Distance Preconditioning''~(DP) technique, first proposed
in~\cite{deDivitiis:2010ya}. Rather than modifying the Dirac operator
directly, as it was done in the original paper by de Divitiis et al., 
we included this preconditioning in the Dirac operator inversion via 
multiplication with a diagonal matrix $P$,
\be
P =
\arraycolsep=10.0pt
\begin{pmatrix}
p_1 & 0 & \ldots & \ldots & 0 \\
0 & p_2 & 0 & \ldots & 0 \\
0 & 0 & \ddots & 0 & 0 \\
0  & \ldots & 0 & \ddots & 0 \\
0 & \ldots & \ldots & 0 & p_T \\
\end{pmatrix},
\ee
with 
$p_i = \exp{}\left( \alpha_0 \cdot \vert{} y_0 - x_{0}^{(i)} \vert{} \right)$,
corresponding to timeslice $i$, and a (control) parameter $\alpha_0$. 
$P$ acts as unity matrix in spin, color and spatial coordinates. 
By this procedure, 
timeslices far away from the source are enhanced by a large exponential
factor, which balances the rapid decay of the correlation function and
ensures an accurate computation of the propagators even in the case of 
large quark masses (in lattice units) and very large time extents.

For our implementation, instead of numerically solving the 
un-preconditioned system
\begin{equation}
A S = \eta{} \qquad \mathrm{with} \qquad A = \sum_y (D[U]+m_0)_{x,y}, 
\label{eq:uncond}
\end{equation}
we multiply it with $P$ from the left and rewrite
$(P A P^{-1}) (P S) = (P \eta)$ as $A' S' = \eta'$. 
The inversion to calculate the solution $S' = PS$ of the preconditioned
system then becomes numerically much more stable, and to recover the 
original solution $S$, it suffices to scale $S'$ back with $P^{-1}$.
\subsection{Numerical tests}
\noindent
We performed several exploratory tests on three different subsets of the 
CLS ensembles H105, H200 and U101. These low-statistic analyses 
(usually 25 to 200 configurations were included in the correlator
measurements) served to explore the validity of the distance preconditioning 
technique for two different values of $\beta$ and different quark masses. 
The U101 subset was chosen because of its larger temporal lattice extent 
of $T/a=127$.

Since there are contaminations from excited states near the source and
excitations close to the boundary, the two-point functions
defined in~\eqref{eq:corrs}~have the general form
\begin{equation} 
 C_{\rm A,P}(x_0,y_0)= 
 A_1(y_0)\,{\rm e}^{-m_{{\rm D}_{\rm q}}(x_0-y_0)}+A_2(y_0)\,{\rm e}^{-m'(x_0-y_0)} 
 +B_1(y_0)\,{\rm e}^{-(E_{2{\rm D}_{\rm q}}-m_{{\rm D}_{\rm q}})(T-x_0+y_0)}+\ldots, 
\end{equation} 
where the first term mimics the ground state, the second one
represents the first excited state with mass $m'$ and the third term
is the first excited boundary state with 
$E_{2{\rm D}_{\rm q}} \approx 2m_{{\rm D}_{\rm q}}$ (see Ref.~\cite{Bruno:2016plf}). 
As one moves farther away from the boundary, the excited states become 
strongly suppressed, and the behaviour of the two-point functions is 
governed by the first term.

\begin{figure}[h!]
  \begin{minipage}{0.5\textwidth}
     \centering
     \includegraphics[width=1.0\textwidth]{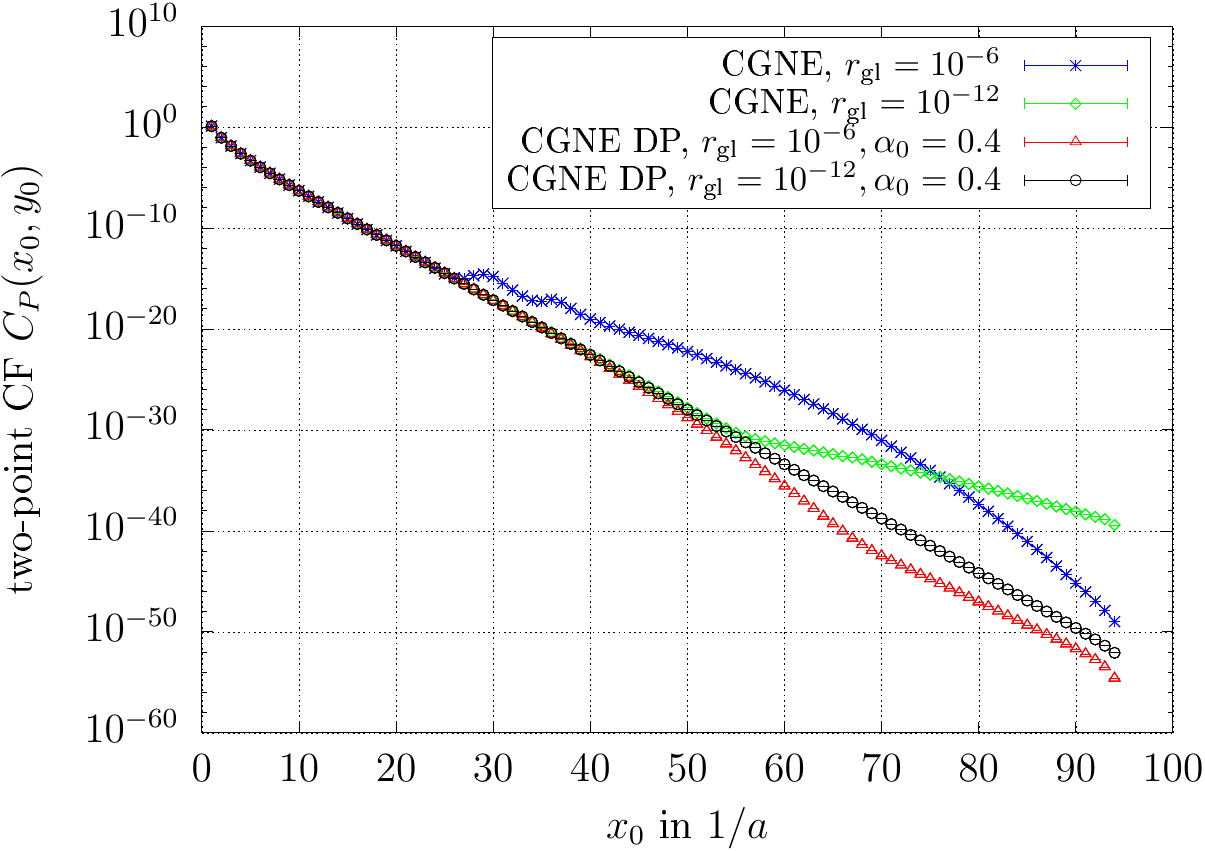}
  \end{minipage}
  \hfill
  \begin{minipage}{0.5\textwidth}
    \centering
    \includegraphics[width=1.0\textwidth]{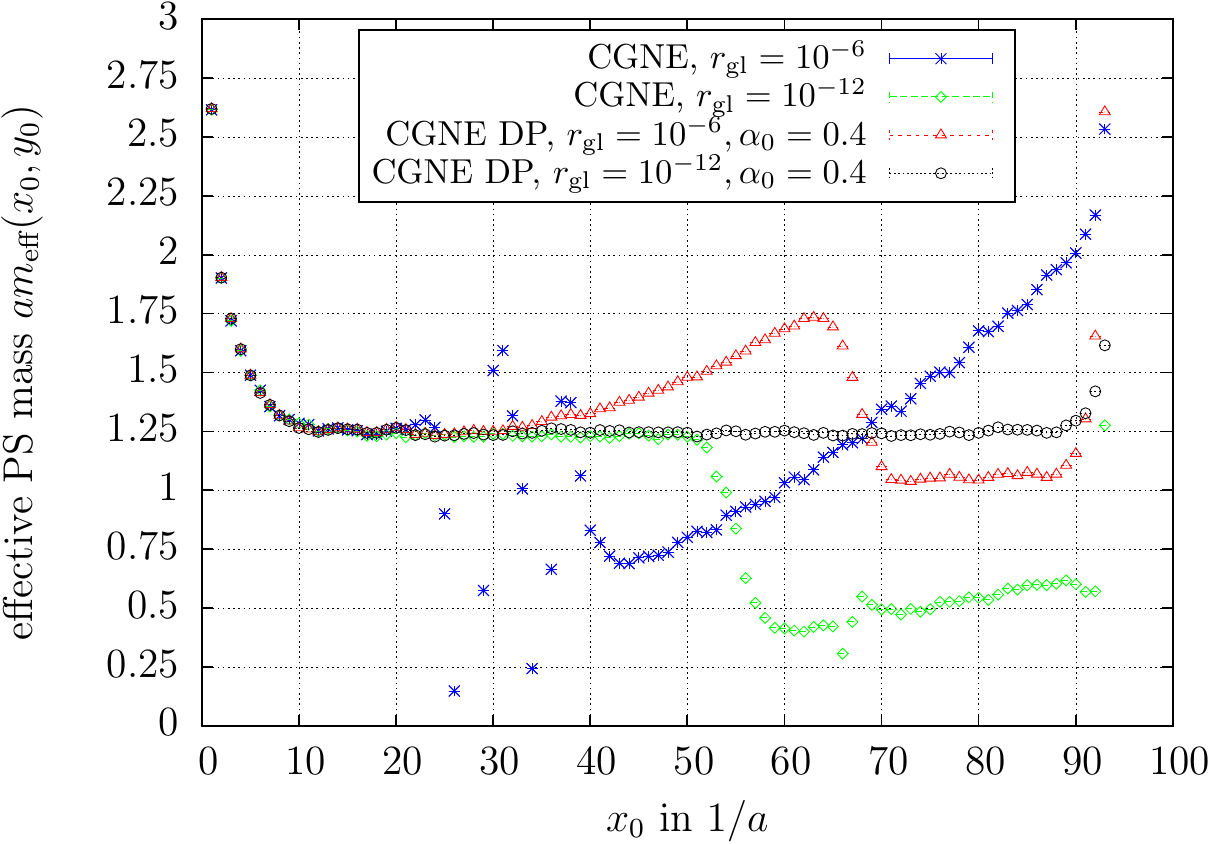}
  \end{minipage}
   \caption{Two-point correlation function $C_{\rm P}(x_0,y_0)$ (left panel) and effective mass 
     $a m_{\text{eff}}(x_0,y_0)$ (right panel) for an exploratory run on the 
     H105 subset. Shown are the results for the unmodified CGNE solver with 
     global residual $r_{\text{gl}}=10^{-6}$ (blue) and $r_{\text{gl}}=10^{-12}$
     (green), as well as those for the DP-modified CGNE solver with 
     $r_{\text{gl}}=10^{-6},\alpha_0=0.4$ (red) and 
     $r_{\text{gl}}=10^{-12},\alpha_0=0.4$ (black).}
   \label{fig:cgne} 
\end{figure}

First tests of the distance preconditioning method were performed
employing the normal-equation variant of the conjugate gradient solver
(CGNE), because of its simplicity and numerical robustness.
Figure~\ref{fig:cgne} illustrates the numerical difficulties, as the
correlator $C_{\rm P}(x_0,y_0)$ deviates from the expected behaviour
(i.e., linear on a logarithmic scale) for large time separations
$\vert x_0 - y_0 \vert$.  These results were obtained on ensemble H105
at $\beta=3.4$ with the correlation function averaged over two source
positions $y_0/a=1$ and $T/a-1$ (DP-results were always generated
utilising these source positions). The heavy-heavy correlator
is evaluated for $\kappa=0.124503\approx\kappa_{\text{charm}}$.  With a
loose global residual of $r_{\text{gl}}=10^{-6}$ the results for the
unmodified CGNE solver (in blue) show an irregular behaviour already
around timeslice $\approx26$. While this can still be partly improved
via choosing a far smaller residual of order $10^{-12}$ (in green),
it is obvious that brute-force methods become unfeasible on lattices
with larger time extent.  The outcome can, however, be greatly
improved by distance preconditioning: While the results for the
DP-modified CGNE solver with $r_{\text{gl}}=10^{-6}$ and control
parameter $\alpha_0=0.4$ (red) already yield a modest improvement,
those for $r_{\text{gl}}=10^{-12}$ (black) show a smooth and nearly
perfect exponential decay over the whole range such that the effective
pseudoscalar meson mass $am_{\text{eff}}(x_0,y_0)= \ln \left(
\frac{C_{\rm P}(x_0,y_0)}{C_{\rm P}(x_0+a,y_0)} \right)$ can be
extracted along a wide plateau (the slight increase in
$am_{\text{eff}}(x_0,y_0)$ at $x_0/a>85$ is likely due to boundary
excited states back-propagating into the bulk).

\begin{figure}[h!]
  \begin{minipage}{0.5\textwidth}
     \centering
     \includegraphics[width=1.0\textwidth]{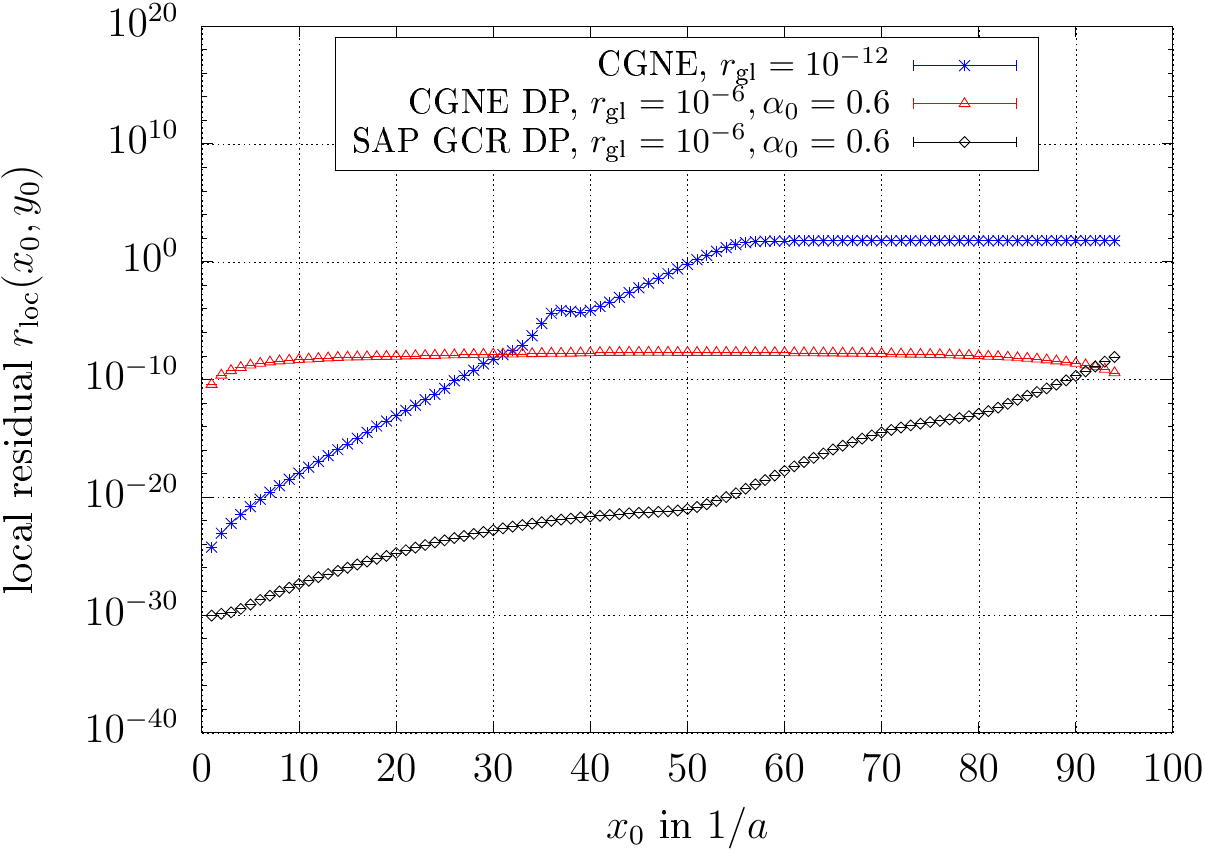}
  \end{minipage}
  \hfill
  \begin{minipage}{0.5\textwidth}
    \centering
    \includegraphics[width=1.0\textwidth]{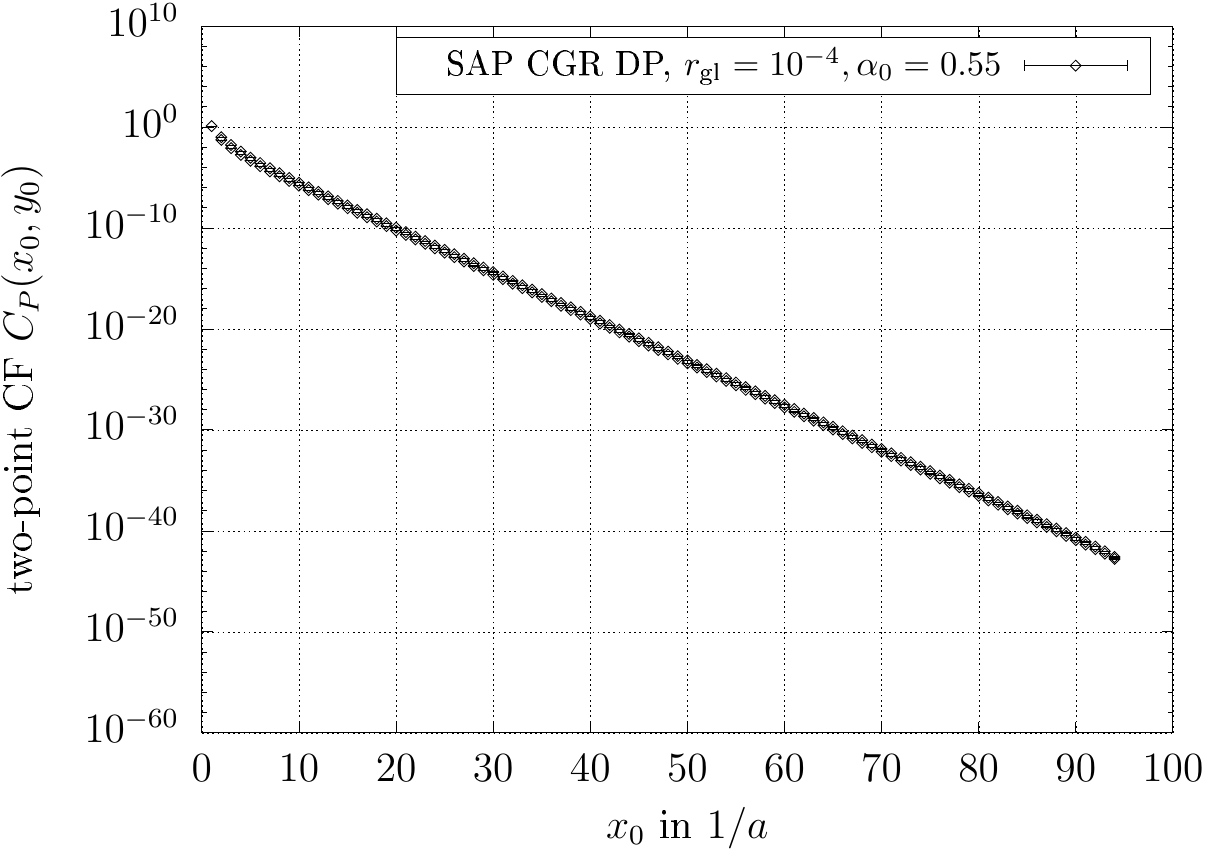}
  \end{minipage}
   \caption{Left panel: Local residual $r_{\text{loc}}(x_0,y_0)$
   plotted against time in lattice units. Shown are the results for
   the unmodified CGNE solver with global residual $r_{\text{gl}}=10^{-12}$ 
   (blue), the DP-modified CGNE solver with 
   $r_{\text{gl}}=10^{-6},\alpha_0=0.6$ (red) and the
   DP-modified SAP GCR solver with $r_{\text{gl}}=10^{-6},\alpha_0=0.6$ (black). 
   Right panel:
   Two-point correlator $C_{\rm P}(x_0,y_0)$ on subset H200 at $\beta=3.55$.
   Solver routine: 
   DP-modified SAP GCR with $r_{\text{gl}}=10^{-4},\alpha_0=0.55$.}
   \label{fig:loc_res} 
\end{figure}

In order to gain in computational speed, we also implemented the DP-method
into another solver routine, namely, the domain-decomposition (SAP)
preconditioned generalized conjugate residual solver (GCR). 
The left panel of Figure~\ref{fig:loc_res} shows the results for the 
\textit{local} residual, defined as 
$r_{\text{loc}}(x_0,y_0)=
\frac{\vert A(x_0,y_0) S(x_0,y_0) 
-\eta_t(x_0,y_0)\vert}{\vert S (x_0,y_0)\vert}$,
for three different solver routines. 
While the CGNE solver (blue) shows substantial deviations between source 
$\eta_t(x_0,y_0)$ and solution $S(x_0,y_0)$, this is neither the case for 
the DP-modified CGNE solver residual (red) that stays almost constant over 
the whole time extent, nor the DP-modified SAP GCR solver one (black), 
which stays well below $r_{\text{loc}}=10^{-8}$. 
The right panel shows a representative result for H200 with 
$\beta=3.55$ and $\kappa_l = \kappa_s = m_{\text{sym}}$. 
If $\kappa_{\text{charm}}$ is chosen close to a value corresponding to
$\frac{1}{2}am_{\text{eff}}^{\text{heavy-heavy}}$, one can afford to set the 
global residual to even larger values, which helps to decrease the computational 
cost of the inversion further.

\begin{figure}[h!]
  \begin{minipage}{0.5\textwidth}
     \centering
     \includegraphics[width=1.0\textwidth]{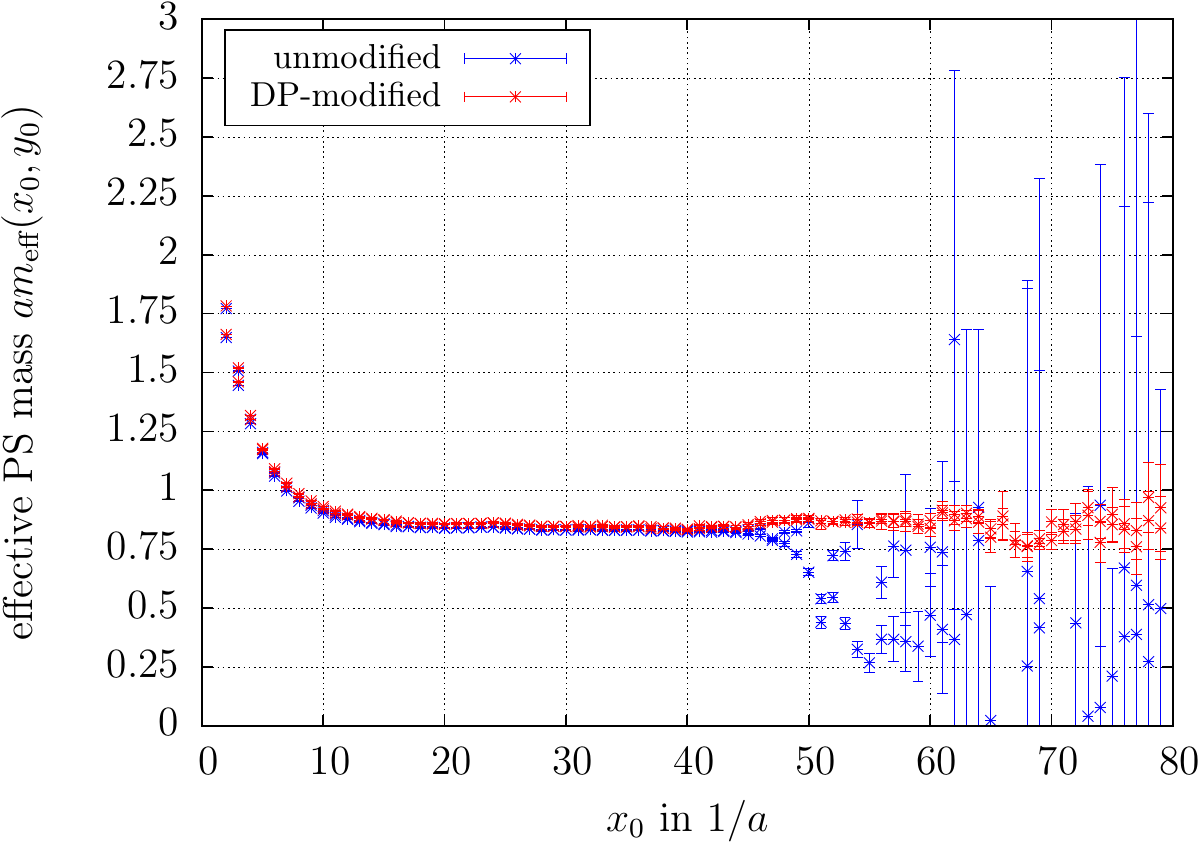}
  \end{minipage}
  \hfill
  \begin{minipage}{0.5\textwidth}
    \centering
    \includegraphics[width=1.0\textwidth]{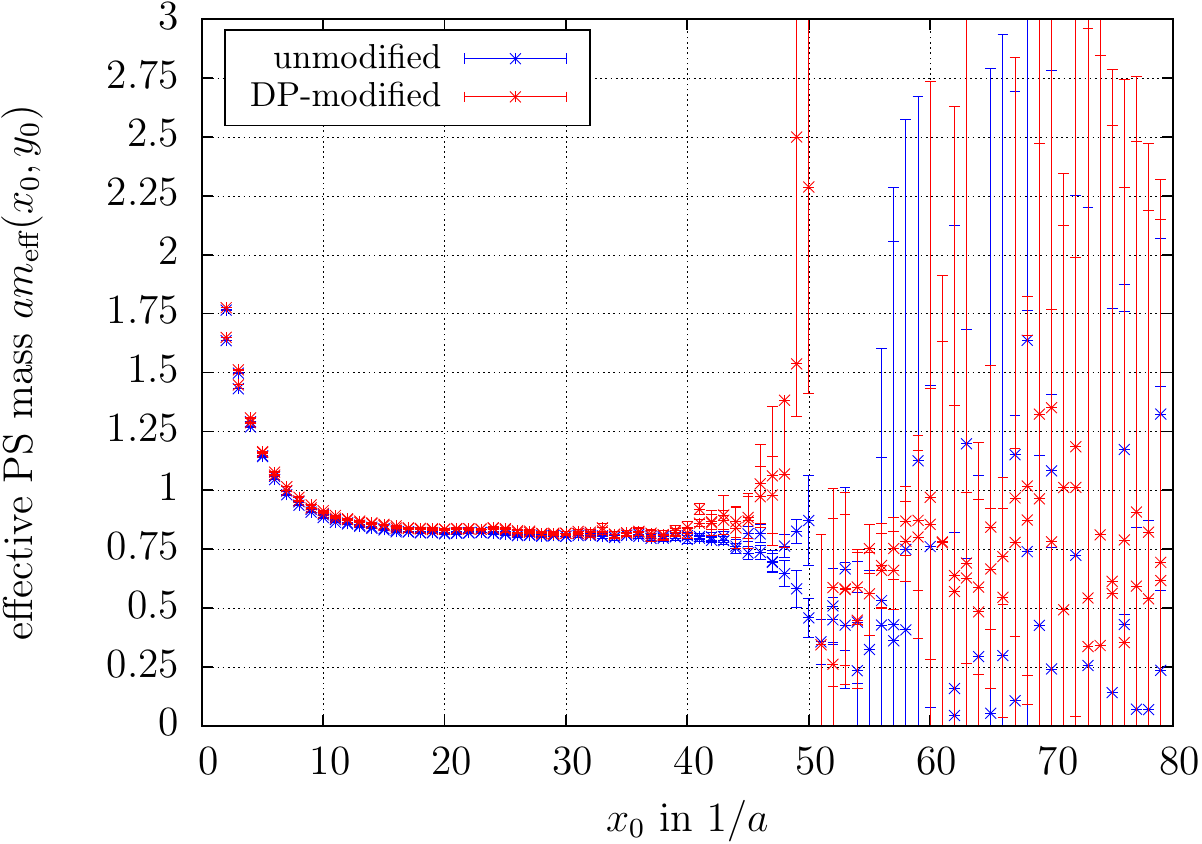}
  \end{minipage}
   \caption{Effective mass $am_{\text{eff}}(x_0,y_0)$ in lattice units
     for heavy-strange 
     (left panel) and heavy-light combinations (right panel); 
     H105, based on 50 configurations per simulation run. 
     The charm channel solutions were obtained with an unmodified 
     DFL SAP GCR solver (blue) with $r_{\text{gl}}=10^{-11}$ and
     the DP-modified SAP GCR solver (red) with 
     $r_{\text{gl}}=10^{-4},\alpha_0=0.7$.
     Solutions for light and strange quarks stem from the unmodified
     DFL SAP GCR solver with $r_{\text{gl}}=10^{-11}$.}
   \label{fig:hshl} 
\end{figure}

Finally, Figure~\ref{fig:hshl} highlights the influence of DP on
$\mathrm{D}_{\rm s}$ meson (left panel) and D meson (right panel)
effective masses.  While the light and strange quark systems were
solved in all cases by a SAP-preconditioned GCR solver with local
deflation (DFL), the charm quark inversion was performed
either in the same way or via the DP-modified SAP GCR solver (red) for
comparison.  In the $\mathrm{D}_{\rm s}$ meson case, an increase in
accuracy is clearly visible, as the plateau region extends farther
into the bulk.  For the D meson, the benefits of the DP-technique are
less pronounced.  Further studies based on higher statistics and
devoted to finding optimal choices for $r_{\text{gl}}$ and $\alpha_0$
are certainly needed, given the promising results of our first
numerical tests.

\section{Analysis details}
\label{sec:ana}
\noindent
Since the tests of the distance preconditioning technique and its
final implementation within our correlator measurement setup are still
on-going, the results we present below were generated by solving the
\textbf{un}-preconditioned linear system~(Eq.~\ref{eq:uncond}) with a
decreased residual~(Eq.~\ref{eq:residual}) of $r_{\rm gl}=10^{-15}$ in
the charm and $r_{\rm gl}=10^{-11}$ in the strange and light quark
propagator computations.
With this choice of (global) residual for the charm
quark inversion, the numerical problems discussed in the previous
section can be seen in charmonium two-point functions at
distances of around 60 timeslices from the source.

As mentioned in Section~\ref{sec:cls}, we use twisted-mass reweighting
for the light quarks and the rational approximation for the strange
quark in the HMC.  To correct to the proper distribution, we must
reweight the physical observables (in this case, the two-point
functions) with the associated reweighting factors:
\begin{equation}
 \langle O \rangle=
 \frac{\langle OW_0W_1\rangle_W}{\langle W_0W_1\rangle_W}.
\end{equation}
The twisted-mass~($W_0$) and rational approximation
reweighting~($W_1$) factors are defined in
Ref.~\cite{Bruno:2014jqa} (Eqs. (3.2) and (3.5), respectively).  

\begin{figure}[t]
 \centerline{
 \includegraphics[width=0.5\textwidth]{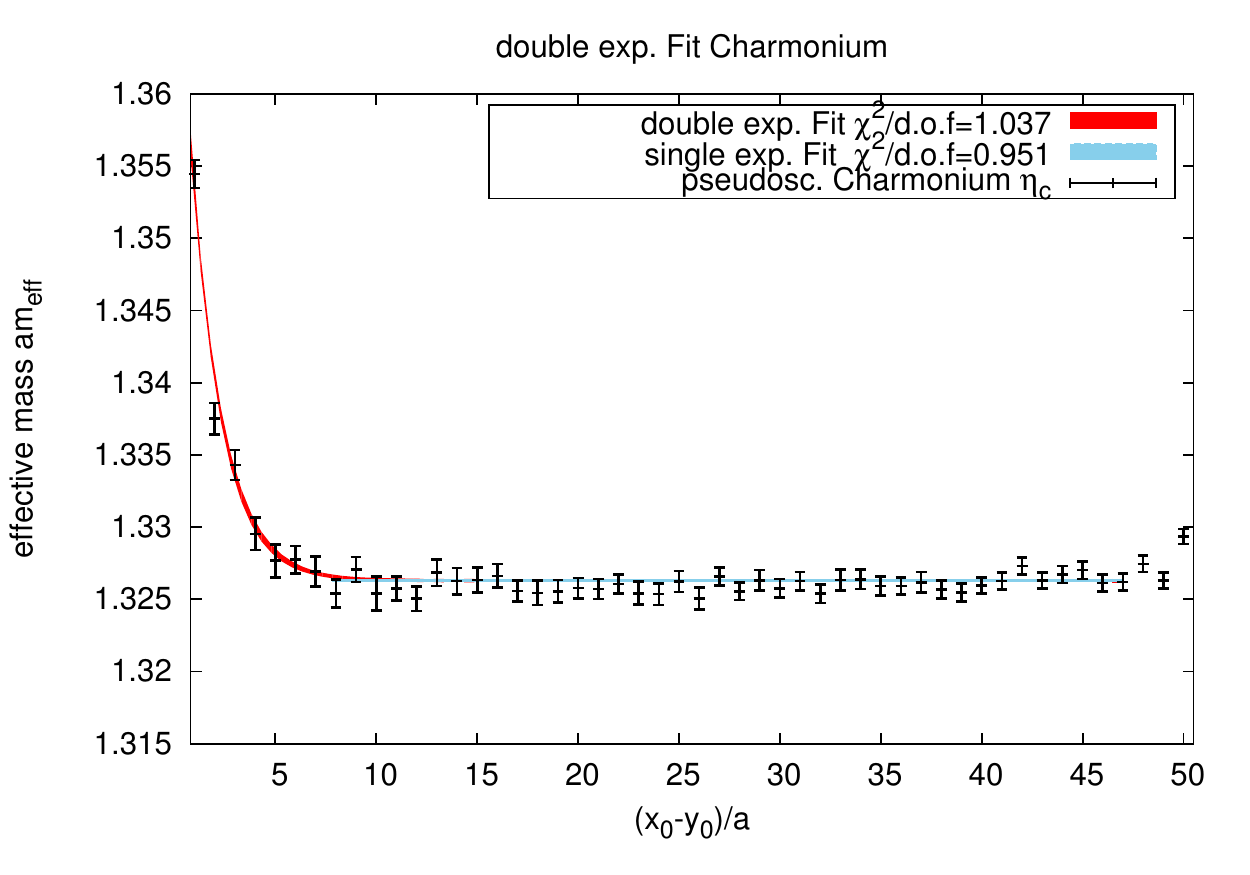}
 \includegraphics[width=0.5\textwidth]{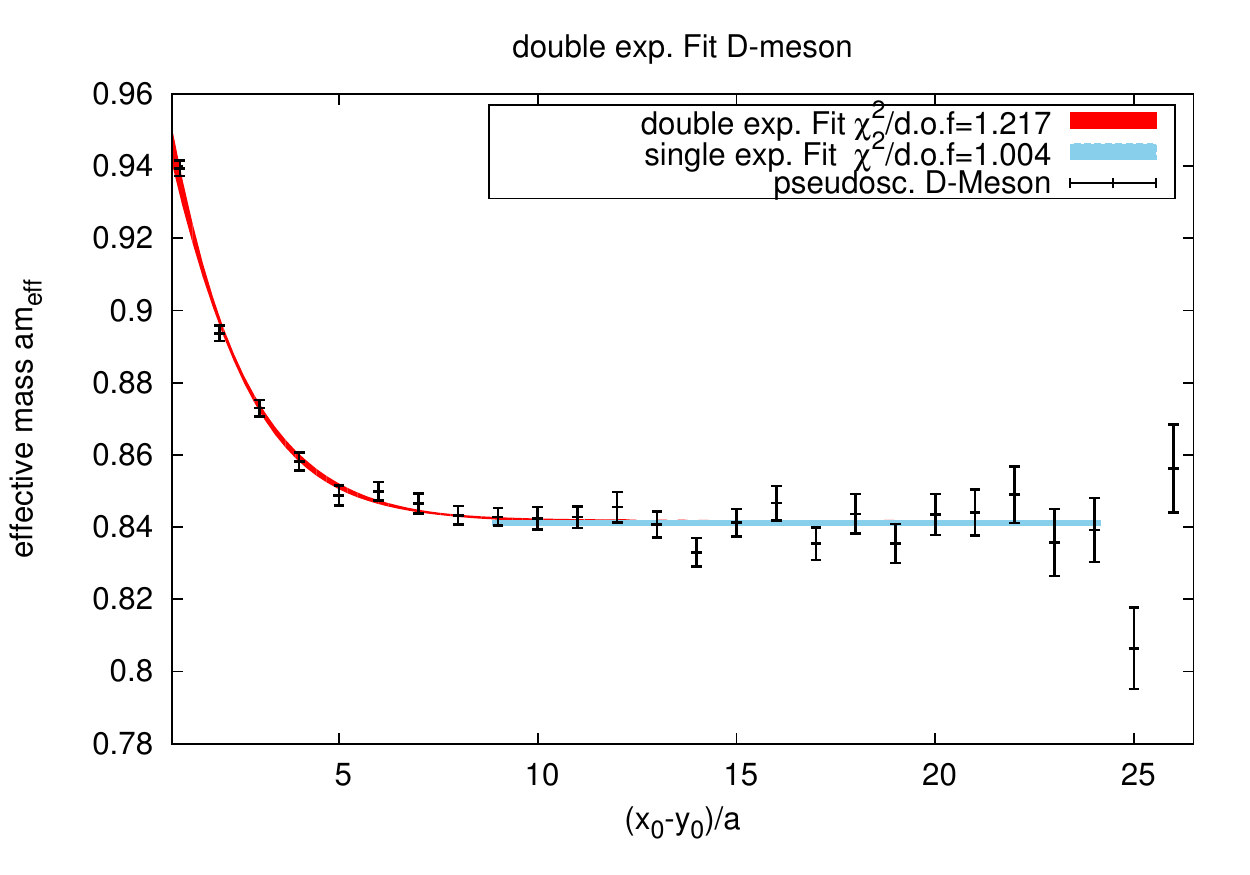}
 }
 \caption{Determination of the fit range: the effective masses of the
   pseudoscalar charmonium (left) and D meson (right) as a function of
   the source-sink separation $(x_0-y_0)/a$ for the H105 ensemble.
   The $C_{\rm P}$ correlators are averaged over 
   $y_0=30a$~(forward-propagating) and $65a$~(backward-propagating),
   where $T/a=95$. 
   A double-exponential fit~(red shaded region) is shown with
   $(x_0^{\rm min},x_0^{\rm max})=(a, 47a)$~(left) and ($a,24a$)~(right).
   These fits are then used to determine the fitting range for the
   single-exponential fits~(see text), indicated by the blue shaded
   regions. }
 \label{fig:Fit}
\end{figure}

The analysis of the two-point functions proceeds by first making use
of time-reversal symmetry and averaging $C_{\rm X}(x_0,y_0)$ and
$C_{\rm X}(T-x_0,T-y_0)$ (${\rm X}={\rm A},{\rm P}$, see
Eq.~\ref{eq:corrs}) for the three source positions, e.g.~$y_0=30a,47a$
and $65a$ for $T/a=95$ at $\beta=3.4$. In other words, we average the
forward-propagating part of $y_0=30a$ with the backward-propagating
branch of $y_0=65a$, while for correlators with $y_0=47a$, we average
the forward and backward propagating mesons.  To extract the lattice
pseudoscalar decay constant, we fit the averaged correlators for
$C_{\rm A}$ and $C_{\rm P}$ simultaneously~(i.e., four correlators in
total) to the single-exponential forms in Eq.~\ref{eq:corrs_exp}. Of
course, this functional form is only valid in a region, where
contamination from excited states has fallen below the noise and
boundary effects as well as the numerical problems associated with the
charm quark inversion are absent.  As discussed in
Section~\ref{sec:obs}, the values of $y_0$ have been chosen to be
sufficiently far away from the boundary.  Effects from the opposite
boundary and the numerical inversion problems are not significant when
fixing the end point of the fit range to $x^{\rm max}_0-y_0\lesssim
25a$ for the D meson and $x^{\rm max}_0-y_0\lesssim 30a$ for the
$\mathrm{D}_\mathrm{s}$ meson.  In order to select the starting point
of the fit interval, $x_0^{\rm min}$, we first estimated the
contribution of the excited states by fitting the correlators to a
double-exponential form.  $x_0^{\rm min}$ is then fixed to the point
at which excited state contributions amount to less than one quarter
of the statistical error of the correlation function.  For example,
for $C_{\rm P}$, \be
\frac{|A^\prime|^2\,\mathrm{e}^{-M^\prime(x_0^{\rm
      min}-y_0)}}{2M^\prime} < \frac{1}{4}\,\Delta C_{\rm P}(x_0^{\rm
  min},y_0), \ee where $A^\prime$ and $M^\prime$ denote the amplitude
and the mass of the first excited state, respectively, as determined
by the fit, and $\Delta C_{\rm P}$ denotes the statistical error of
the correlator.  Figure~\ref{fig:Fit} illustrates this procedure.  The
lattice pseudoscalar decay constant is then extracted from a
combination of the amplitudes of $C_{\rm A}$ and $C_{\rm P}$
(resulting from single-exponential fits between $x_0^{\rm min}$ and
$x_0^{\rm max}$), in accordance with Eq.~\ref{eq:corrs_exp}.

As the charm quark is partially quenched in our setup, we need to fix
the corresponding hopping parameter~($\kappa_{\rm charm}$) in our
analysis.  In a first step, on a subset of the available statistics,
$\kappa_{\rm charm}$ was estimated from the charmonium 1S
spin-averaged mass $\overline{M}(1S)=\left(m_{\eta_c}
+3m_{J/\Psi}\right)/4$.  In order to maintain flexibility for possible
adjustments later, the correlation functions were calculated with full
statistics at two values of the hopping parameter, straddling the
target $\kappa_{\rm charm}$; these values were chosen by allowing the
lattice spacing to vary by $\pm 2\%$.  $\kappa_{\rm charm}$ was then
determined on this full set by interpolating $\overline{M}(1S)$
linearly in the inverse of the hopping parameter to the physical point,
as given by the central value for $a$.  A linear interpolation is valid,
since both $\kappa$-values for the ``heavier'' and ``lighter'' charm
quark are close enough to the physical value.  Employing the
charmonium 1S mass introduces an unknown~(although expected to be
small) uncertainty, since we neglect the disconnected diagrams and
possible flavour mixing.  As an alternative, free from these
systematics, we also determined $\kappa_{\rm charm}$ utilising the
spin-flavour-averaged 1S meson mass combination
$M_X= \left(6m_{{\rm D}^*}+2m_{\rm D}+3m_{{\rm D}_{\rm s}^*}
+m_{{\rm D}_{\rm s}}\right)/12$ along the $\overline{m}=\mathrm{const.}$
line, and the spin-averaged ${\rm D}_{\rm s}$ mass along the 
$\hat{m}_{\rm s}=\mathrm{const.}$ line.
So far, the results for the interpolated
decay constants, using $\kappa_{\rm charm}$ fixed via either
$\overline{M}(1S)$ or $M_X$ and $m_{{\rm D}_{\rm s}}$, were consistent
within one standard deviation.  In the future, we plan to base our
final determination of the hopping parameter of the (valence) charm
quark on $M_X$ and $m_{{\rm D}_{\rm s}}$.

\section{Preliminary results}
\label{sec:res}
\begin{figure}[htb]
\centerline{%
\includegraphics[width=0.5\textwidth]{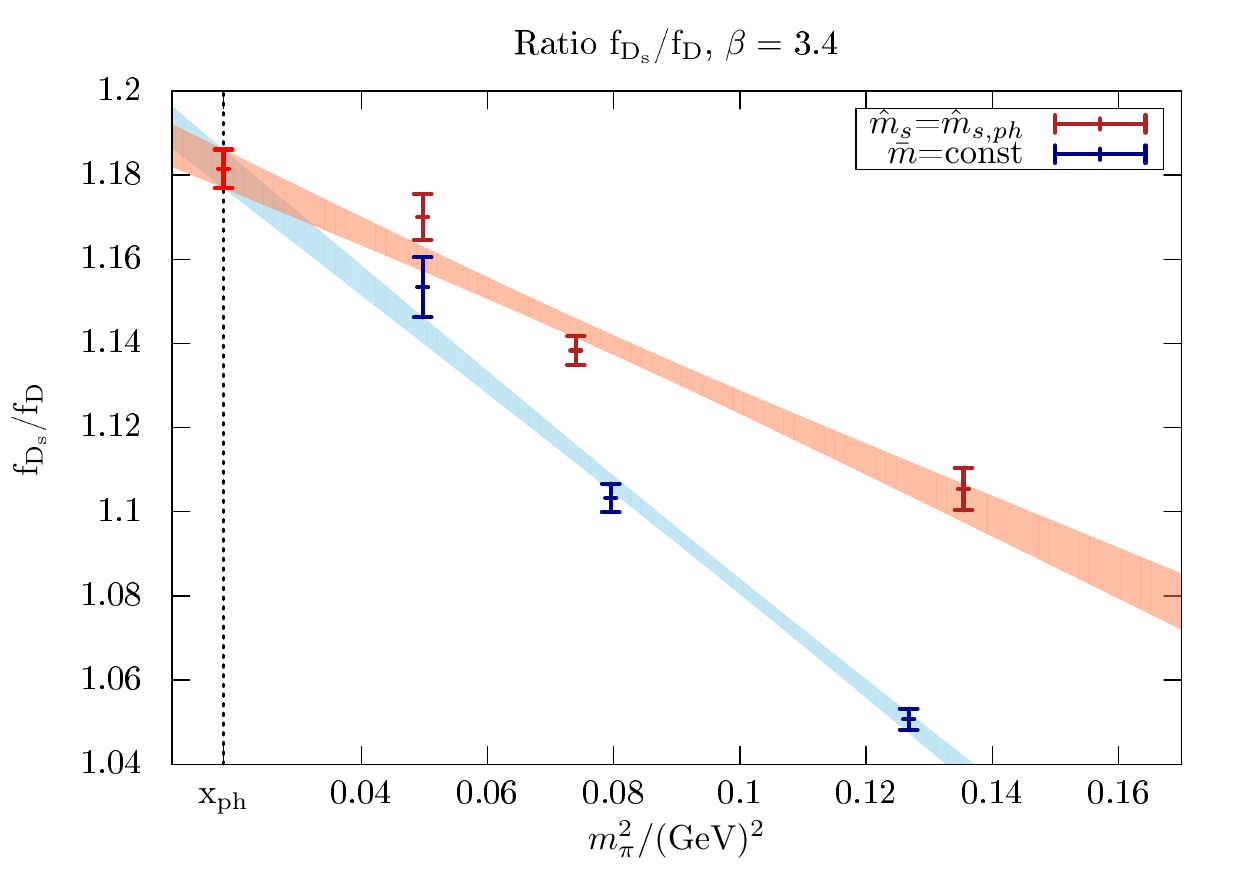}
\includegraphics[width=0.5\textwidth]{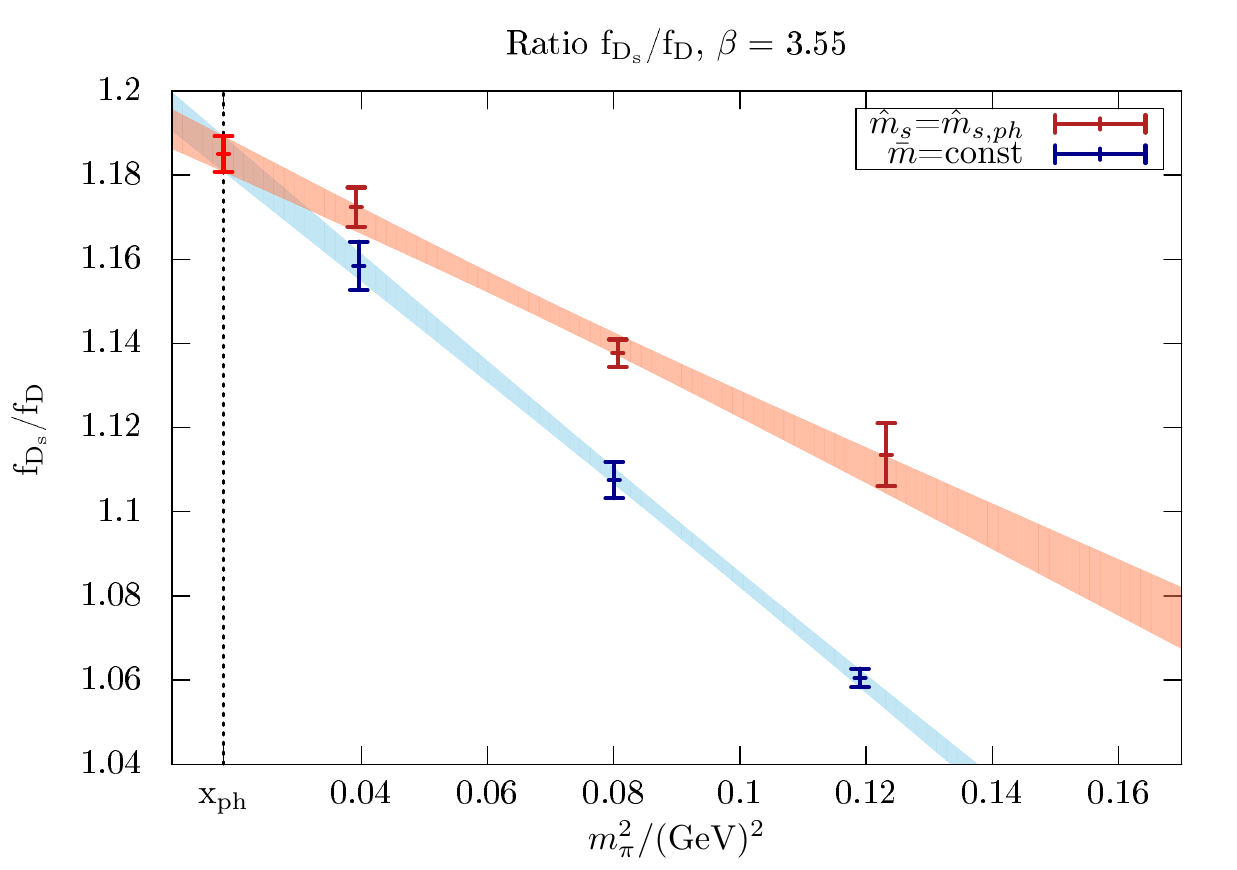}}
    \caption{Ratio $\fDs/\fD$
      for $\beta$=3.4 (left panel) and $\beta$=3.55 (right panel). 
      The blue points refer to data from ensembles along
      the $\overline{m}={\rm const.}$ line, whereas red points symbolize
      data along the $\ms={\rm const.}$ trajectory. We perform a
      linear chiral extrapolation simultaneously along both
      trajectories (orange- and blue-shaded regions), enforcing the 
      value at the physical point~(dashed vertical line) to be the same.}
    \label{fig:FDsoverFD}
\end{figure}

\begin{figure}[htb]
\centerline{%
\includegraphics[width=0.5\textwidth]{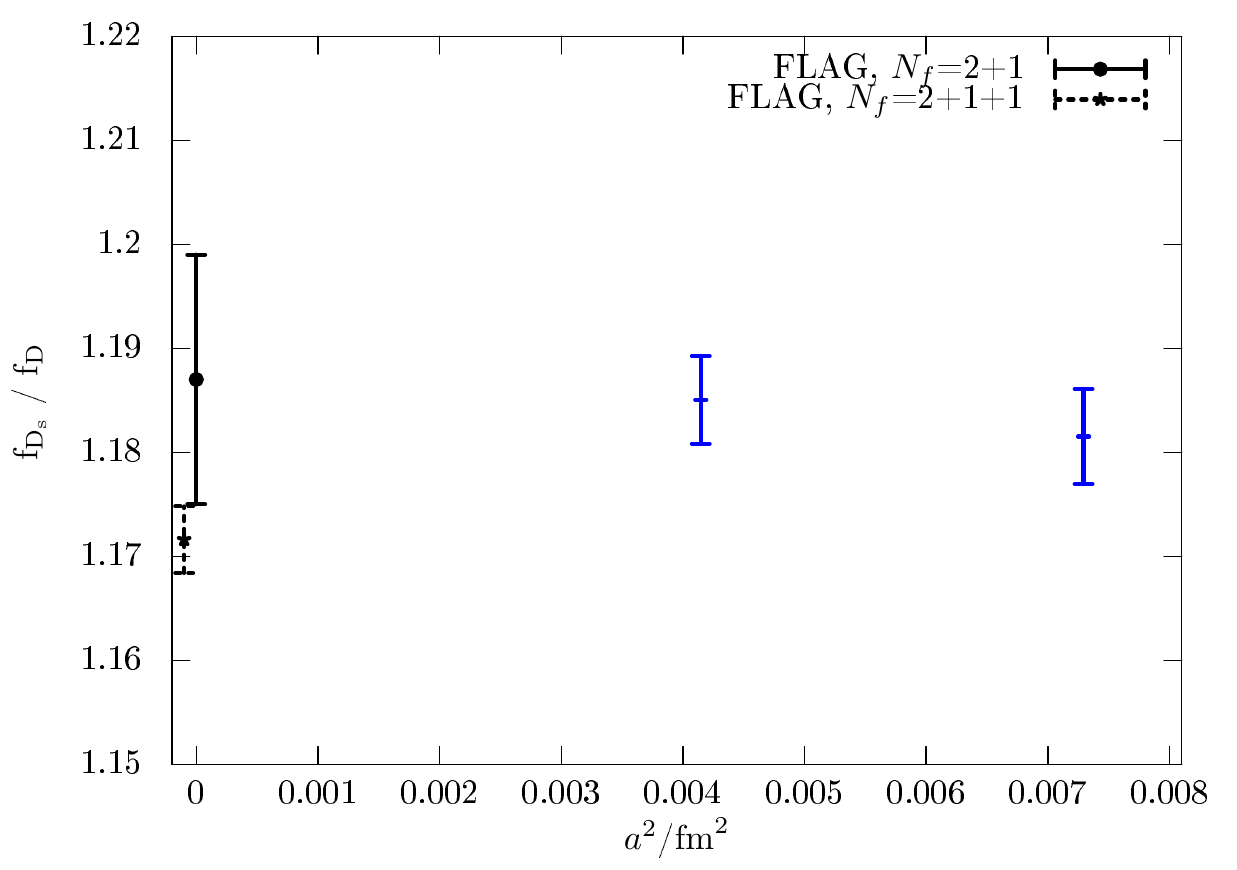}}
    \caption{Our preliminary results for $\fDs/\fD$ for $\beta$=3.4
      and 3.55 as a function of $a^2$, compared to the recent averages
      for $\nf=2+1$ and $\nf=2+1+1$ from
      FLAG~\protect\cite{Aoki:2016frl}. Note that the errors on our
      results are only statistical at the moment.  }
    \label{fig:cont}
\end{figure}

\noindent
The results presented in this section are preliminary and systematic
errors arising from the uncertainty in the lattice spacing, fixing the
charm quark mass, the renormalisation procedure, etc. are neglected.
The statistical errors are estimated using a bootstrap analysis
employing 500 samples.
Let us first inspect the ratio of decay constants, $\fDs/\fD$, for which 
the renormalisation factor~($\za$) drops out and also the contributions 
from the ${\cal O}(a)$ improvement terms 
partially cancel~(see Eqs.~(\ref{eq:CaImprove}) and~(\ref{eq:RenImp})).
Figure~\ref{fig:FDsoverFD} shows our results for the ratio as a
function of $m_\pi^2$ for $\beta=3.4$ and 3.55~(corresponding to
$a=0.0854\,\Fm$ and $0.0644\,\Fm$, respectively), consult 
Table~\ref{tab:CLSensembles} for the values of $m_\pi$ for each ensemble.
We performed chiral extrapolations linear in $m_\pi^2$ along the
trajectories of constant $\overline{m}$ and 
$\hat{m}_{\rm s}=\hat{m}_{\rm s}^{\rm phys}$ simultaneously, enforcing the 
value at the physical point to be the same in both cases. 
As the dominant discretisation artefacts are of ${\cal O}(a^2)$, 
we show the results for the two $\beta$-values plotted against $a^2$ in 
Figure~\ref{fig:cont}. The comparison with the recent FLAG 
averages for $\nf=2+1$ and $\nf=2+1+1$ \cite{Aoki:2016frl} is promising.  

\begin{figure}[h!]
\centerline{%
\includegraphics[width=0.5\textwidth]{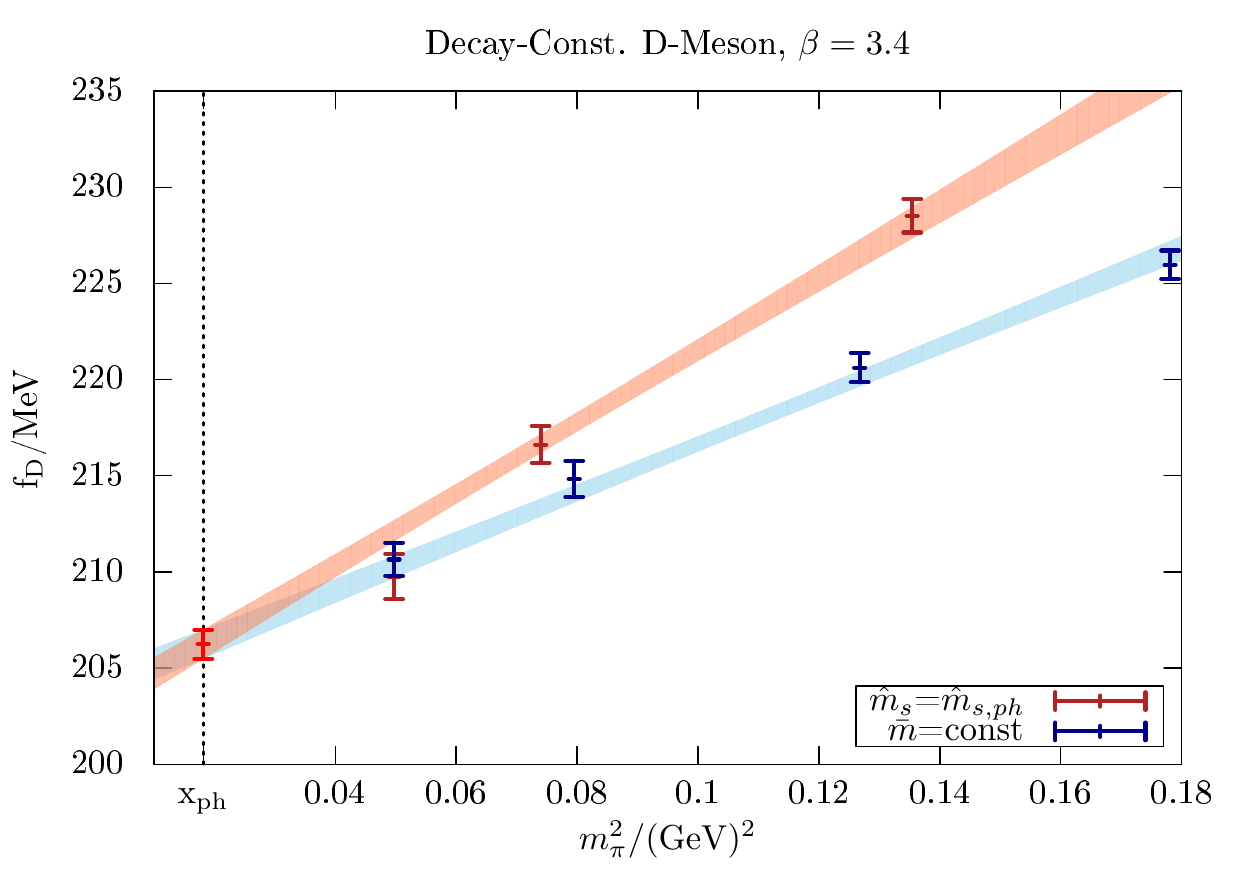}
\includegraphics[width=0.5\textwidth]{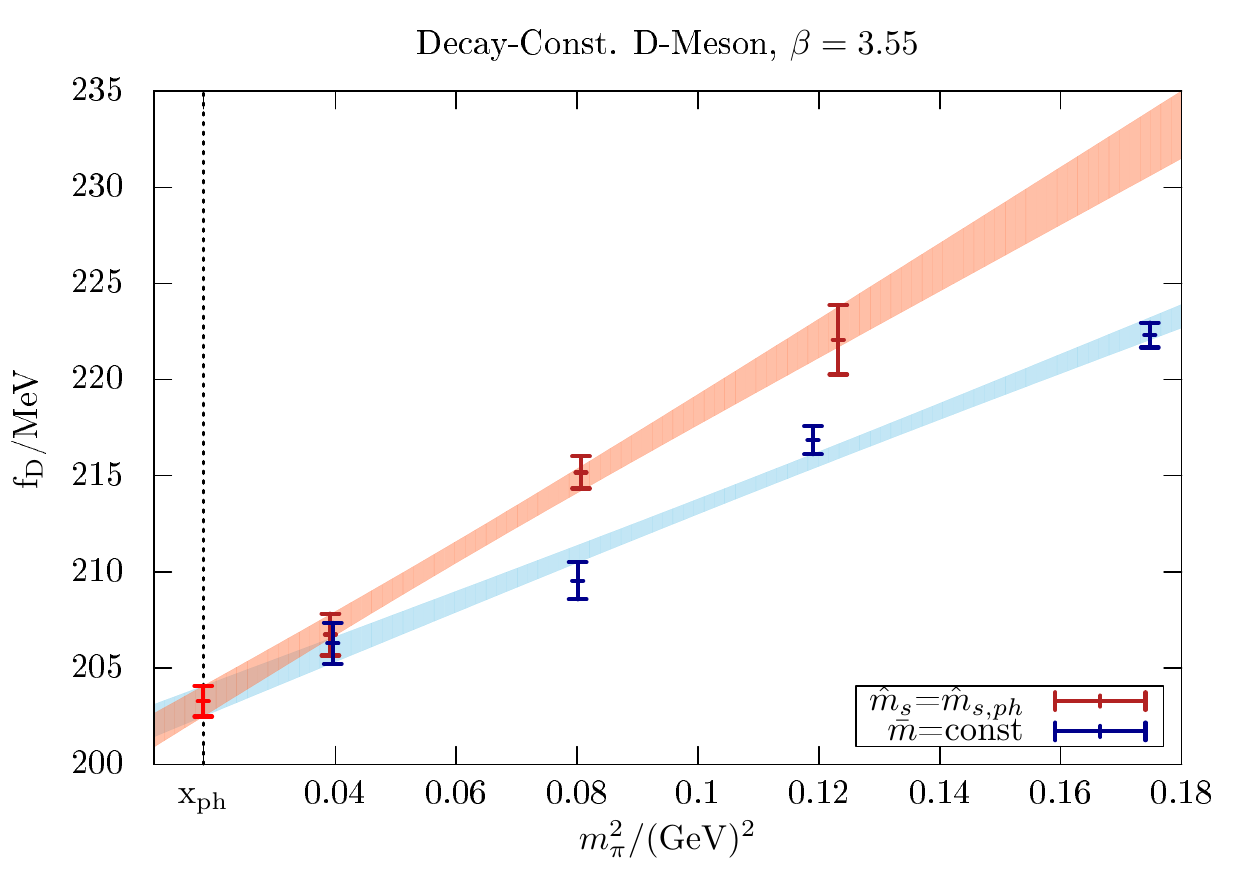}}
    \caption{$\fD$ for $\beta$=3.4 (left panel) and $\beta$=3.55 
      (right panel) as in Figure~\protect\ref{fig:FDsoverFD}.}
    \label{fig:FD}
\end{figure}

\begin{figure}[h!]
\centerline{%
\includegraphics[width=0.5\textwidth]{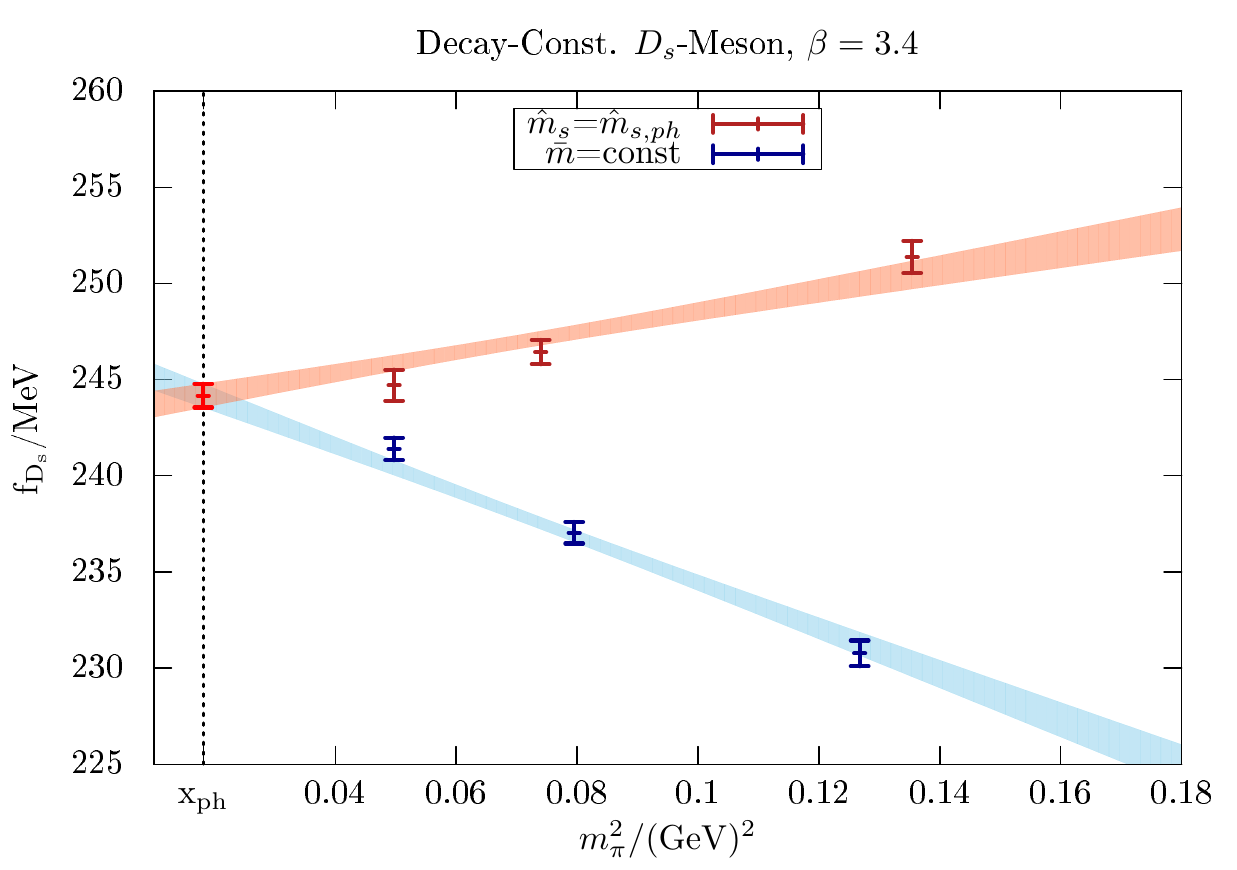}
\includegraphics[width=0.5\textwidth]{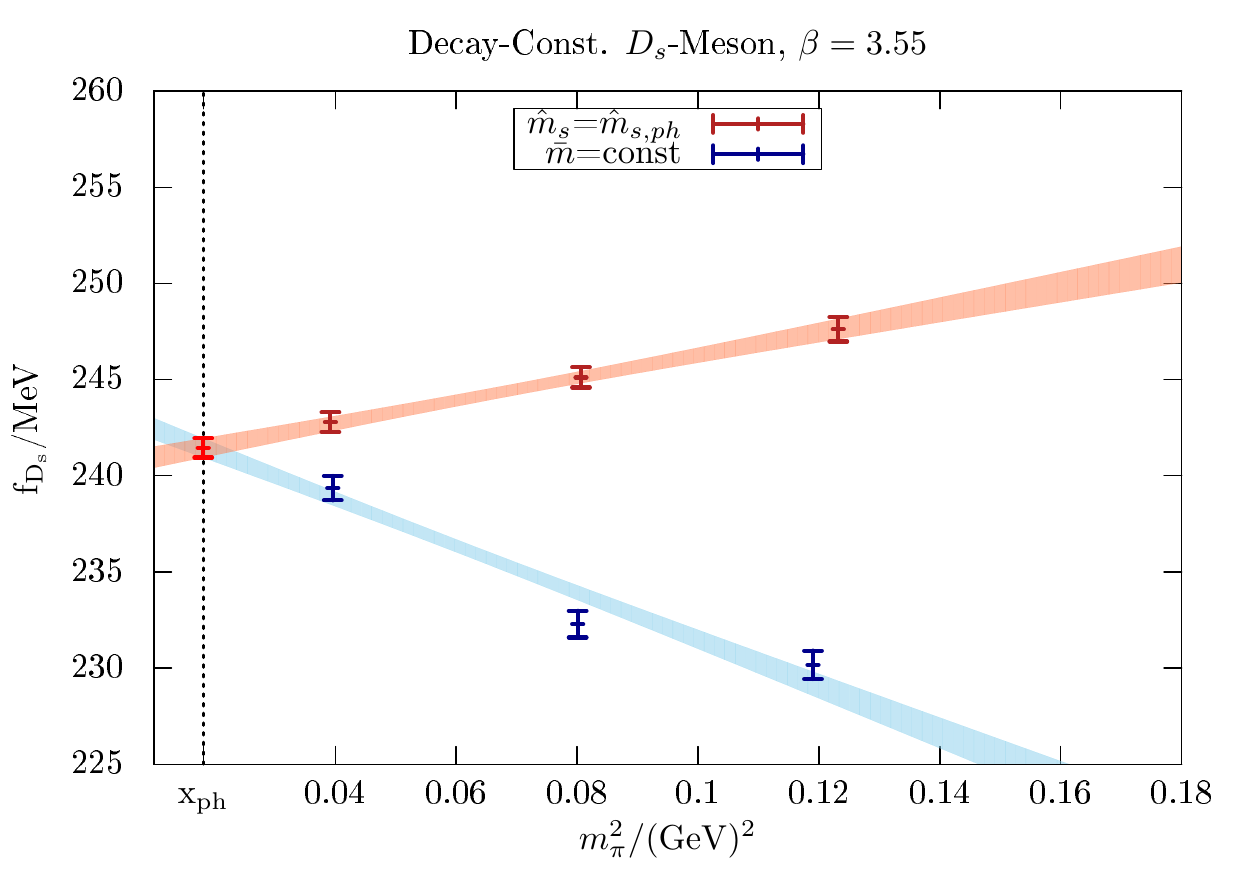}}
    \caption{$\fDs$ for $\beta$=3.4 (left panel) and $\beta$=3.55 
      (right panel) as in Figure~\protect\ref{fig:FDsoverFD}.}
    \label{fig:FDs}
\end{figure}

The individual pseudoscalar D and ${\rm D}_{\rm s}$ meson decay
constants are displayed in Figures \ref{fig:FD} and~\ref{fig:FDs}.
These depend on the renormalisation, and ${\cal O}(a)$ improvement has
a big impact.  If we focus on the $\mathrm{D}_{\mathrm{s}}$ meson,
e.g., including the $\ba m_{\rm qc}$-improvement term in
Eq.~(\ref{eq:RenImp}) causes an upward shift of the physical-point
value of $\fDs$ of about 35\% for $\beta=3.4$ and 19\% on the finer
$\beta=3.55$ lattice.  Similar shifts occur for $\fD$.  Most of the
shift is due to the presence of the charm quark mass in $m_{\rm qc}$.
In contrast, for the ratio $\fDs/\fD$ there is only a 1\% shift when
adding this improvement term, in line with the expectation that the
impact of the charm quark largely cancels here.  In the future, the
chiral extrapolations will be investigated in more detail and, once
results at additional lattice spacings have been analysed,
a careful continuum limit extrapolation will also be performed.


\section{Conclusions and outlook}
\label{sec:concl}
\noindent
Our ongoing simulations in the charmed meson sector of $\nf=2+1$
lattice QCD address the computation of the leptonic decay constants
$\fD$ and $\fDs$.  To reach a precision high enough to be relevant for
theory inputs to global analyses in flavour physics
phenomenology, among the next necessary steps are: (i)~increase of
statistics, inclusion of further ensembles and correcting the results
for possible small mismatches of the constant physics condition, 
enabling a stable joint chiral and continuum extrapolation;
(ii)~a full budget of (statistical and systematic) errors, also
accounting for their correlations.
This will be carried out in the future.


\vspace{0.5cm}
{\footnotesize%
\noindent {\bf Acknowledgments.}
We thank Gunnar Bali, Tomasz Korzec, Stefan Schaefer, Rainer Sommer
and Nazario Tantalo for useful discussions.
This work is supported by the Deutsche Forschungsgemeinschaft (DFG)
through the grants GRK~2149
(\emph{Research Training Group 
``Strong and Weak Interactions -- from Hadrons to Dark Matter''},
K.~E. and J.~H.) and SFB/TRR~55 (S.~C., S.~H. and W.~S.). 
We are indebted to our colleagues in CLS for the joint production of
the $\nf=2+1$ gauge configurations.
The authors gratefully acknowledge the Gauss Centre for 
Supercomputing e.V. for granting computer time on SuperMUC at the 
Leibniz Supercomputing Centre. 
Additional simulations were performed on the Regensburg iDataCool
cluster and on the SFB/TRR~55 QPACE~2 and QPACE~B
computers~\cite{Baier:2009yq,Nakamura:2011cd}.
Some of the two-point functions were computed using the
Chroma~\cite{Edwards:2004sx} software package, along with the locally
deflated domain decomposition solver implementation of
openQCD~\cite{openQCD}, the LibHadronAnalysis library and the
multigrid solver implementation of Ref.~\cite{Heybrock:2015kpy};
our implementation of distance preconditioning (Section~\ref{sec:dp}) 
is based on~\cite{mesons}.
}

%
%
\bibliography{lat16pap_nf3fDs}{}
\bibliographystyle{JHEP-2_notitles}
%
%
\end{document}